\documentclass[prl,aps,twocolumn,preprintnumbers,superscriptaddress,showpacs,nofootinbib]{revtex4-2}
\usepackage[colorlinks=true, pdfstartview=FitV, linkcolor=red, citecolor=blue, urlcolor=blue, pdftitle={},pdfsubject={}, pdfkeywords={}]{hyperref}

\newcommand\sect[1]{{\it #1.}---}

\usepackage{stmaryrd}
\usepackage{comment}
\usepackage{amsmath}
\usepackage{amssymb}
\usepackage{amsthm}
\usepackage{mathrsfs}
\usepackage{graphicx}
\usepackage{marvosym}
\usepackage{amsmath,bm}
\usepackage{array}
\usepackage{latexsym}
\usepackage{enumerate}
\usepackage{slashed}
\usepackage{hyperref}
\usepackage{color}

\allowdisplaybreaks[1]

\usepackage[normalem]{ulem}

\begin{document}
\title{Local spin polarization by color-field correlators and momentum anisotropy}
\author{Haesom Sung}
\email{ioussom@gmail.com }
\affiliation{
	Institute of Physics, Academia Sinica, Taipei, 11529, Taiwan.\\
}
\author{Berndt M\"uller}
\email{bmueller@duke.edu}
\affiliation{
Department of Physics, Duke University, Durham, North Carolina 27708, USA.\\
}
\author{Di-Lun Yang}
\email{dilunyang@gmail.com}
\affiliation{
Institute of Physics, Academia Sinica, Taipei, 11529, Taiwan.\\
}
\affiliation{Physics Division, National Center for Theoretical Sciences, Taipei, 106319, Taiwan}
\begin{abstract}
We study the local spin polarization of quarks induced by color-field correlators stemming from the correlation of chromo-Lorentz force and chromo-magnetic polarization or chromo-spin Hall effect in the presence of momentum anisotropy. 
Such effects can trigger longitudinal polarization from fluctuating color fields in glasma or quark gluon plasma phases with transverse expansion for relativistic heavy ion collisions. Especially, from the glasma effect, the resulting longitudinal polarization spectrum of $\Lambda/\bar{\Lambda}$ hyperons has a sinusoidal structure with twice the azimuthal angle relative to the anisotropic direction. An order-of-magnitude estimate of the effect aligns with experimental observations. Our findings highlight the significant role of coherent gluon fields as a novel source for spin polarization phenomena in high-energy nuclear collisions.
\end{abstract}

\maketitle

\sect{Introduction}
Recent experimental findings on the global spin polarization of $\Lambda/\bar{\Lambda}$ hyperons in relativistic heavy ion collisions (HIC) \cite{STAR:2017ckg,STAR:2018gyt} have prompted theoretical investigations into the genesis and transport of quark spin polarization in quark-gluon plasmas (QGP). A widely accepted hypothesis suggests that the substantial angular momentum in peripheral collisions leads to the spin polarization of quarks and gluons via spin-orbit coupling, the polarization subsequently being transferred to hadrons during hadronization and freeze-out \cite{Liang:2004ph,Liang:2004xn}. While global thermal equilibrium of hadrons or constituent quarks primarily links their polarization to thermal vorticity \cite{Becattini2013a,Fang:2016vpj,Becattini:2013vja}, additional modifications arise under conditions of local equilibrium \cite{Hidaka:2017auj,Liu:2020dxg,Liu:2021uhn,Becattini:2021suc}. 

Particularly, the thermal shear mechanism has been identified as a significant factor in compensating for thermal vorticity, thereby contributing to the generation of longitudinal spin polarization (LSP) along the beam direction \cite{Fu:2021pok,Becattini:2021iol,Yi:2021ryh,Ryu:2021lnx,Florkowski:2021xvy,Sahoo:2024egx}. This effect yields a sinusoidal angular dependence, specifically with twice the azimuthal angle relative to the reaction plane, which is in qualitative agreement with experimental observations \cite{STAR:2019erd}. Nevertheless, the overall sign of LSP, which arises from competing effects, is sensitive to the approximations and input parameters employed. See also Refs.~\cite{Liu:2019krs,Sun:2024isb} for alternative approaches to address the issue. 
 
Furthermore, recent measurements of LSP in proton-nucleus collisions with high-multiplicity events have revealed a new discrepancy between theoretical predictions \cite{Yi:2024kwu} and experimental data \cite{CMS:2025nqr}, underscoring the need for incorporating additional effects to fully understand local spin polarization. Despite extensive theoretical endeavors to investigate non-equilibrium effects \cite{Lin:2022tma,Fang:2022ttm, Fang:2023bbw, Lin:2024zik, Fang:2024vds, Lin:2024svh, Wang:2024lis, Fang:2025pzy, Weickgenannt:2022zxs, Weickgenannt:2022qvh, Weickgenannt:2024ibf, Wagner:2024fry, Sapna:2025yss} through quantum kinetic theory (QKT) \cite{Stephanov:2012ki,Son:2012wh,Chen:2012ca,Hidaka:2016yjf,Gao:2019znl,Weickgenannt:2019dks,Hattori:2019ahi,Wang:2019moi,Li:2019qkf,Yang:2020hri,Weickgenannt:2020aaf,Wang:2020pej,Hattori:2020gqh,Hidaka:2022dmn} and hydrodynamics for spinning fluids \cite{Montenegro:2017rbu,Florkowski:2017ruc,Florkowski:2018fap,Yang:2018lew,Hattori:2019lfp,Fukushima:2020ucl,Shi:2020htn,Li:2020eon,Hongo:2021ona,Bhadury:2022ulr}, the underlying mechanisms that can account for the subtle angular structure of LSP with a relevant order of magnitude, beyond the forms in local equilibrium polarization, remain elusive.

On the other hand, the experimentally observed spin alignment for vector mesons in HIC \cite{ALICE:2019aid,STAR:2022fan} might be induced by the spin correlation of coalesced quarks and antiquarks produced by strong-force fields such as the fluctuating vector-meson fields \cite{Sheng:2022wsy,Sheng:2022ffb,Sheng:2023urn} or glasma fields and color fields from low-energy thermal gluons in QGP \cite{Kumar:2022ylt,Kumar:2023ghs,Yang:2024qpy}. One can ask under what conditions such gluonic contributions could influence the polarization of quarks in connection to the LSP of $\Lambda/\bar{\Lambda}$ hyperons, which can reveal the microscopic features of QCD matter created from HIC.    

In this letter, we employ QKT for massive quarks in the presence of background color fields \cite{Muller:2021hpe,Yang:2021fea} to derive the spin polarization spectrum. This spectrum is induced by color-field correlators, with the momentum anisotropy being characterized by the quark flow velocity. These color-field correlators originate from the intricate correlations involving the chromo-Lorentz force, chromo-magnetic polarization, or the chromo-spin Hall effect, without necessitating the assumption of local equilibrium for spin degrees of freedom. In fact, a similar phenomenon was reported in Refs.~\cite{Muller:2021hpe,Yang:2021fea}, where spin polarization could only be generated by the parity-odd correlator between chromo-electric and chromo-magnetic fields in the absence of anisotropic flow. 

We demonstrate in this Letter that parity-even correlators, arising from pairs of chromo-magnetic or chromo-electric fields, also induce polarization in the presence of collective flow. We show that this mechanism naturally gives rise to the sinusoidal angular structure observed in LSP. We further estimate the order of magnitude for LSP from color fields led by glasma and QGP phases where the fields are, respectively, longitudinally dominant or isotropic. The effects from the two phases are found to be competing with each other due to the subtlety of the freeze-out hypersurface. Finally, we discuss the significance of these findings particularly in the context of high-multiplicity proton-nucleus collisions.

Throughout this Letter, we use 
the mostly minus signature of the Minkowski metric $\eta^{\mu\nu} = {\rm diag} (1, -1,-1,-1)  $ 
and the completely antisymmetric tensor $ \epsilon^{\mu\nu\rho\lambda} $ with $ \epsilon^{0123} = 1 $. 
We use the notations $A^{(\mu}B^{\nu)}\equiv A^{\mu}B^{\nu}+A^{\nu}B^{\mu}$ 
and $A^{[\mu}B^{\nu]}\equiv A^{\mu}B^{\nu}-A^{\nu}B^{\mu}$ and also define $\tilde{F}^{a\mu\nu}\equiv\epsilon^{\mu\nu\alpha\beta}F^{a}_{\alpha\beta}/2$ with $F^{a\mu\nu}$ being the field strength of color fields and the upper index $a$ representing color. The chromo-electric and chromo-magnetic fields are explicitly given by $F^{a}_{\mu\nu}=-\epsilon_{\mu\nu\alpha\beta}B^{a\alpha}\bar{n}^{\beta}+E^a_{[\mu}\bar{n}_{\nu]}$
, where $\bar{n}^{\mu}=(1,\bm 0)$ denotes the temporal direction.

\sect{Spin polarization from color-field correlators}
Before introducing QKT, we provide an intuitive argument for generating spin polarization from color fields. We consider the chromo-Lorentz force, ${\bm F}^a=g({\bm E}^a+{\bm u}\times{{\bm B}^a})$, and the color-octet polarization pseudo-vector from chromo-magnetic polarization and chromo-spin Hall effect, ${\bm P}^a=g({\bm B}^a-{\bm u}\times{{\bm E}^a})$, for the quark transport, where ${\bm u}$ as a flow velocity delineates momentum anisotropy. The color-singlet polarization pseudo-vector for quarks with momentum ${\bm p}$ can accordingly be constructed by the correlation, $\langle ({\bm p}\cdot{\bm F}^a) {\bm P}^a\rangle\sim ({\bm p}\times {\bm u})(\langle {\bm B}^{a}\cdot{\bm B}^{a}\rangle+\langle {\bm E}^{a}\cdot{\bm E}^{a}\rangle)$, when having just non-vanishing parity-even color-field correlators. Here we have introduced $\langle\,\rangle$ to denote the ensemble average over color sources. We now derive this effect in the QKT.

We start with an outline for essential steps in the derivation of the spin polarization spectrum of quarks with background color fields constructed in Ref.~\cite{Muller:2021hpe,Yang:2021fea}. In general, the building blocks of the spin polarization spectrum are the vector and axial-vector components of Wigner functions for massive fermions, $\mathcal{V}^{\mu}(p,x)$ and $\mathcal{A}^{\mu}(p,x)$, responsible for the particle-number and spin-current density in phase space, respectively. For quarks carrying color charges, they can be further decomposed into the color-singlet and color-octet components, 
\begin{eqnarray}\nonumber
	\mathcal{V}^{\mu}(p, x)&=&\mathcal{V}^{{\rm s}\mu}(p, x)I+\mathcal{V}^{a\mu}(p, x)\,t^a,
	\\
	\mathcal{A}^{\mu}(p, x)&=&\mathcal{A}^{{\rm s}\mu}(p, x)I+\mathcal{A}^{a\mu}(p, x)\,t^a,
\end{eqnarray}
where $t^a$ are the SU$(N_c)$ generators such that $[t^a,t^b]=if^{abc}t^c$ and $\{t^a,t^b\}=N_c^{-1}\delta^{ab}I+d^{abc}t^c$ and $I$ is the identity matrix in color space.
Assuming the spin of $\Lambda$ hyperons is primarily dictated by the spin of strange quark therein (and same for the $\bar{\Lambda}$ and strange antiquark) and the perfect transition of the polarization from quarks to hadrons, known as the strange-equilibrium scenario \cite{Fu:2021pok,Yi:2021ryh}, the polarization spectrum of $\Lambda$ hyperons is given by   
\begin{eqnarray}
	\mathcal{P}^{\mu}({\bm p})=\frac{{\rm tr}_{\rm c}\big[\int d\Sigma\cdot p\mathcal{A}^{\mu}(\bm p, x)\big]}{2M_{
		\Lambda}{\rm tr}_{\rm c}\big[\int d\Sigma\cdot \mathcal{V}({\bm p},x)\big]},
\end{eqnarray}
where $\mathcal{A}^{\mu}(\bm p, x)\equiv(2\pi)^{-1}\int dp_0\mathcal{A}^{\mu}(p, x)$ and $\mathcal{V}^{\mu}(\bm p, x)\equiv(2\pi)^{-1}\int dp_0\mathcal{V}^{\mu}(p, x)$ as the onshell Wigner functions and $d\Sigma^{\mu}$ denotes the normal vector of a spin freeze-out hypersurface. Here $M_{\Lambda}$ corresponds to the mass of $\Lambda$ hyperons in light of the phenomenological choice for normalization. Since ${\rm tr}_{\rm c}$ as the trace over color space is taken, we will focus on the color-singlet components contributing to $\mathcal{P}^{\mu}({\bm p})$.

Based on the quantum nature of spin, we adopt the power counting, $\mathcal{V}^{\mu}\sim\mathcal{O}(\hbar^0)$ and $\mathcal{A}^{\mu}\sim\mathcal{O}(\hbar)$ in the $\hbar$ expansion as the gradient expansion in phase space \cite{Yang:2020hri,Hidaka:2022dmn}, and consider the leading-order contributions led by
\begin{eqnarray}
\mathcal{V}^{{\rm s}\mu}({\bm p},x)=\left(\frac{p^{\mu}}{2p_0}f^{\rm s}_{V}(p,x)\right)_{p_0=\epsilon_{\bm p}}
\end{eqnarray}
and
\begin{eqnarray}\nonumber\label{eq:As_onshell}
\mathcal{A}^{{\rm s}\mu}(\bm p, x)&=&\frac{1}{2\epsilon_{\bm p}}\Big[\tilde{a}^{{\rm s}\mu}(p,x)-\frac{\hbar g}{4N_c}\tilde{F}^{a\mu\nu}(x)\partial_{p\nu}f^{a}_{V}(p,x)\Big]_{p_0=\epsilon_{\bm p}}
\\
&&+\frac{\hbar g}{8N_c}\tilde{F}^{a\mu\nu}(x)\partial_{p_\perp\nu}\big(f^{a}_{V}(p,x)/\epsilon_{\bm p}\big)_{p_0=\epsilon_{\bm p}},
\end{eqnarray}
where $\epsilon_{\bm p}\equiv\sqrt{{\bm p}^2+m^2}$ and $p_{\perp\nu}\equiv p_{\nu}-p_0\bar{n}_{\nu}$. We also introduce $\partial_{V\mu}\equiv \partial/\partial V^{\mu}$ and $\partial^{\mu}_{V}\equiv \partial/\partial V_{\mu}$. Here $f^{\rm s}_{V}(p,x)$ and $f^{a}_{V}(p,x)$ denote the color-singlet and color-octet vector-charge distribution functions, respectively. Analogously, $\tilde{a}^{{\rm s}\mu}(p,x)$ represents the color-singlet effective spin four-vector in phase space, while its color-octet counterpart is of higher orders in $\mathcal{V}^{{\rm s}\mu}({\bm p},x)$ and neglected here. In a QCD medium, $f^{a}_{V}(p,x)$ originates from color fluctuations induced by color fields and given $f^{\rm s}_{V}(p,x)$. The $\tilde{F}^{a\mu\nu}$ terms in Eq.~(\ref{eq:As_onshell}) are responsible for chromo-magnetic polarization and the chromo-spin Hall effect.   

The dynamical evolution of the vector-charge distribution functions and effective spin four-vectors are governed by kinetic equations. For weakly coupled systems with slow-varying color fields, dynamical polarization led by non-vanishing $\tilde{a}^{{\rm s}\mu}(p,x)$ is suppressed and we will simply focus on the non-dynamical polarization from the remaining terms of Eq.~(\ref{eq:As_onshell}). At weak coupling, $f^{a}_{V}(p,x)$ can be solved from the collisionless color-octet Vlasov equation perturbatively as the quark evolution under chromo-Lorentz force, 
\begin{eqnarray}\label{eq:fV_octet_sol}
	f_{V}^a(p,x)=-\frac{g}{p_0}\int^{x_0}_{t_{\rm i}} dx'_0p^{\mu}F^{a}_{\nu\mu}(x')\partial_{p}^{\nu}f^{\rm s}_V(p,x')\Big|_{\text{c}},
\end{eqnarray}
where $|_{\text{c}}=\{x^{\prime i}_{\rm T}=x^{i}_{\rm T},x^{\prime i}_{\parallel}=x^{i}_{\parallel}-p^{i}(x_0-x'_0)/p_0\}$ with $x_0$ being the present time and $t_{\rm i}\leq x_0$ denotes an initial time when color fields affect the quark transport.  The subscripts, ${\rm T}$ and $\parallel$, are utilized to represent the components transverse and parallel to $\bm p$. One hence finds
\begin{eqnarray}\nonumber\label{eq:Asmu_FF}
	\mathcal{A}^{{\rm s}\mu}(\bm p, x)&=&\bigg[\frac{\hbar g^2}{8p_0N_c}\int^{x_0}_{t_{\rm i}} dx'_0\langle\tilde{F}^{a\mu\nu}(x)F^{a}_{\alpha\beta}(x')\rangle
	\\
	&&\times \mathcal{D}_{p\nu}\Big(p^{\beta}\partial_{p}^{\alpha}f^{\rm s}_V(p,x')\Big)\Big|_{\text{c}}
	\bigg]_{p_0=\epsilon_{\bm p}},
\end{eqnarray}
where 
\begin{eqnarray}
\mathcal{D}_{p\nu}\equiv\partial_{p\nu}p_0^{-1}-\epsilon_{\bm p}\partial_{p_\perp\nu}\epsilon_{\bm p}^{-2}+p_{\perp\nu}\epsilon_{\bm p}^{-2}\partial_{p0}.
\end{eqnarray}
For a more detailed derivation we refer to Refs.~\cite{Muller:2021hpe,Yang:2021fea}.
 
Standard phenomenology of relativistic HIC suggests that the quarks undergo transverse expansion and their momentum spectrum depends on $p\cdot u/\Lambda_c$, where $u^{\mu}$ is a local collective velocity field and $\Lambda_c$ is a typical scale of the medium. In the early collision stage this transverse flow is thought to be generated by glasma fields \cite{Carrington:2024utf}; in the QGP stage it is governed by hydrodynamics.

When taking $t_{\rm i}$ close to $x_0$ and assuming slowly varying color-field correlators, 
Eq.~(\ref{eq:Asmu_FF}) can be approximated as
\begin{eqnarray}\nonumber\label{eq:Asmu_FF_approx}
	\mathcal{A}^{{\rm s}\mu}(\bm p, x)&\approx&\frac{\hbar g^2\Delta t}{8\epsilon_{\bm p}N_c}\langle\tilde{F}^{a\mu\nu}(x)F^{a\alpha\beta}(x)\rangle u_{\alpha}
	\Big(\hat{O}^{(2)}_{\nu\beta}\partial^2_{p\cdot u}
	\\
	&&+\frac{1}{\epsilon_{\bm p}}\hat{O}^{(1)}_{\nu\beta}\partial_{p\cdot u}\Big)f^{\rm s}_V\left(\frac{p\cdot u}{\Lambda_c}\right)_{p_0=\epsilon_{\bm p}},
\end{eqnarray}
where $\hat{O}^{(2)}_{\nu\beta}=u_0p_{\nu}p_{\beta}/\epsilon_{\bm p}^2$, $\hat{O}^{(1)}_{\nu\beta}=p_{(\nu}\bar{n}_{\beta)}/\epsilon_{\bm p}-2p_{\nu}p_{\beta}/\epsilon_{\bm p}^2$ and $\Delta t=x_0-t_{\rm i}$. As indicated earlier, we here consider only parity-even correlators for paired chromo-electric fields and chromo-magnetic fields: $\langle E^{i}(x)E^{j}(x)\rangle=\delta^{ij}\langle E^{i}(x)E^{i}(x)\rangle$, $\langle B^{i}(x)B^{j}(x)\rangle=\delta^{ij}\langle B^{i}(x)B^{i}(x)\rangle$, and $\langle E^{i}(x)B^{j}(x)\rangle=0$. Selecting $z$ as the longitudinal direction,  Eq.~(\ref{eq:Asmu_FF_approx}) then reduces to 
\begin{eqnarray}\label{eq:Asmu_BB_EE}
	\mathcal{A}^{{\rm s}z}(\bm p, x)&\approx&\frac{\hbar g^2\Delta t}{8\epsilon_{\bm p}^2N_c}\big(\langle B^{a}_{z}(x)B^a_{z}(x)\rangle+\langle E^{a}_{\rm T}(x)E^a_{\rm T}(x)\rangle\big)
	\\\nonumber
	&&\epsilon^{zjk}p^ju^k \big(u^0\partial_{p\cdot u}-\epsilon_{\bm p}^{-1}\big)\partial_{p\cdot u}f^{\rm s}_V\left(\frac{p\cdot u}{\Lambda_c}\right)\Big|_{p_0=\epsilon_{\bm p}},
\end{eqnarray}
where we have 
imposed rotational symmetry on the correlator of transverse color fields $E^{a}_{\rm T}$. Notwithstanding the weak coupling assumption, our derivation should capture the generic form of $\mathcal{A}^{{\rm s}z}(\bm p, x)$ up to linear order in the color-field correlators. Equation (\ref{eq:Asmu_BB_EE}) manifests the angular structure, $\mathcal{\bm P}^z\sim ({\bm u}\times {\bm p})^z$, for LSP, whereas the overall sign will also depends on the freeze-out hypersurface.

\begin{figure}
	\begin{center}
		\includegraphics[width=0.5\hsize]{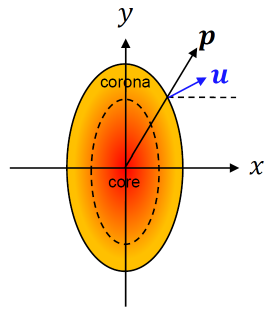}
	\end{center}
	\caption{A schematic figure for the transverse plane of the QCD medium (glasma or QGP), where the region inside the dashed ellipse represents the core, while the region between the solid ellipse and the dashed one corresponds to the corona. Also, $\bm u$ and $\bm p$ denote the flow velocity and momentum of the quark on the medium surface. 
	}
	\label{fig:core_corona}
\end{figure}

\sect{Polarization from glasma} In early times of high-energy nuclear collisions, the overpopulated gluons may be described by classical chromo-electromagnetic fields in the color glass condensate effective theory~\cite{McLerran:1993ni,McLerran:1993ka,McLerran:1994vd,Gelis:2010nm,Albacete:2014fwa} and form the so-called glasma phase~\cite{Lappi:2006fp,Lappi:2006hq}. We may now employ Eq.~(\ref{eq:Asmu_BB_EE}) to study the LSP induced by glasma fields with an initial flow. 
To make a more systematic estimate of the polarization spectrum, we shall take into account the integral over a freeze-out hypersurface. We consider the isochronous freeze-out in 3+1 dimensions at fixed $\tilde{\tau}=\tilde{\tau}_{f}$ with $\tilde{\tau}\equiv\sqrt{\tau^2-x^2-y^2}=\sqrt{t^2-x^2-y^2-z^2}$.
To further delineate the elliptical asymmetry of the transverse plane, we parameterize 
\begin{eqnarray}
	x=r\sqrt{1-\epsilon}\cos\Phi,\quad y=r\sqrt{1+\epsilon}\sin\Phi,
\end{eqnarray}
where $\epsilon$ as a model parameter represents the elongation along the $y$ axis and $r\leq r_{m}$ characterizes the transverse size of the medium.
The four-momentum is parameterized as
\begin{eqnarray}
	p^{\mu}=\big(\epsilon_{T}\cosh y_{p},p_{T}\cos\phi, p_{T}\sin\phi, \epsilon_{T}\sinh y_{p}\big),
\end{eqnarray} 
where $\epsilon_{T}\equiv\sqrt{m^2+p_{T}^2}$ and $y_{p}$ denotes the momentum rapidity. Based on e.g., Refs.~\cite{Schenke:2010nt,Kumar:2023ojl}, it is found
\begin{eqnarray}
d\Sigma\cdot p=drd\Phi d\eta \,r\sqrt{1-\epsilon^2}\big(G_1\cosh(y_p-\eta)+G_2\big),
\end{eqnarray}
where
\begin{eqnarray}
	G_1=\sqrt{(m^2+p_{T}^2)}\tau_{f},\quad G_2=-(xp^x+yp^y),
\end{eqnarray}
with $\tau_f(x,y)=\sqrt{\tilde{\tau}_{f}^2+x^2+y^2}$. Without the loss of generality for semi-quantitative estimates, we may parameterize the four velocity as
\begin{eqnarray}
	u^{\mu}=\frac{1}{N_{v}}(1,\, v^x,\, v^y,\,0),
\end{eqnarray}
where $N_{v}=\sqrt{1-v_x^2-v_y^2}$ with
\begin{eqnarray}
v^x= u_{T}(1+\delta)\frac{x\tau}{r_m^2},\quad
v^y= u_{T}(1-\delta)\frac{y\tau}{r_m^2}.
\end{eqnarray}
Here $u_{T}$ and $\delta$ are model parameters characterizing the strength and anisotropy of flow. Such initial flow may be induced by pressure gradients of glasma \cite{Chen:2013ksa,Chen:2015wia}. 

We first focus on the contribution from the corona of the glasma as depicted in Fig.~(\ref{fig:core_corona}), for which the quarks may immediately hadronize after the glasma phase without thermalizing as opposed to the quarks from the core that will further go through the QGP phase before hadronization. In the glasma phase, the longitudinal color fields are supposed to be more dominant. We hence omit the correlator from transverse chromo-electric fields and neglect the time evolution of color fields in the corona. To make an order-of-magnitude estimate, we adopt the Golec-Biernat W\"usthoff (GBW) dipole distribution  \cite{Golec-Biernat:1998zce,Guerrero-Rodriguez:2021ask} to approximate
\begin{eqnarray}
	g^2\langle B^{az}(t_{\rm i})B^{az}(t_{\rm i})\rangle\approx\frac{(N_c^2-1)}{2N_c}Q_s^4,
\end{eqnarray}
where $Q_s$ denotes the saturation scale.
We postulate two types of quark distributions in glasma. One characterizes the Schwinger pair production by chromo-electric fields in early times\footnote{The original pair production rate is given by $\Gamma\sim e^{-\pi m^2/|eE|}$ \cite{Schwinger:1951nm}. Here we simply make the generalization to replace $m$ by $p\cdot u$ to capture the momentum anisotropy of produced quarks and $|eE|$ by $|gE^a|$. See Refs.~\cite{Glendenning:1983qq,Kerman:1985tj,Gatoff:1987uf,Tanji:2010eu,Taya:2016ovo} for follow-up studies of the Schwinger mechanism by chromo-electric fields.}, 
\begin{eqnarray}\label{eq:quark_Sch_dis}
	f^{\rm s}_V(p\cdot u/Q_s)=e^{-\pi (p\cdot u)^2/|gE^a|} 
\end{eqnarray}
with $|gE^a|\approx \sqrt{(N_c^2-1)/(2N_c)}Q_s^2$ based on the GBW distribution.
Another is a quasi-equilibrium form for quarks at relatively late times in the glasma,
\begin{eqnarray}\label{eq:quark_dis}
f^{\rm s}_V(p\cdot u/Q_s)=\frac{1}{e^{p\cdot u/Q_s}+c},
\end{eqnarray}
eventually approaching thermal equilibrium characterized by $Q_{s}\rightarrow T$ and $c\rightarrow 1$. Here we set $0\leq c\leq 1$ to describe the overpopulation of quarks driven by the high density of gluons at low energy while satisfying the Pauli exclusion principle. We then evaluate the LSP via
\begin{eqnarray}\label{eq:Pz_glasma_origin}\nonumber
	\mathcal{P}^z&\approx& \frac{\hbar (N_c^2-1)Q_s^2\Delta t}{16M_{\Lambda}N_c^2\epsilon_{\bm p}\int d\Sigma\cdot pf^{\rm s}_V}\int d\Sigma\cdot p(p^xu^y-p^yu^x)
	\\
	&&\times \big(u^0\partial_{p\cdot u/Q_s}-Q_s/\epsilon_{\bm p}\big)\partial_{p\cdot u/Q_s}f^{\rm s}_V.
\end{eqnarray}  

\begin{figure}[t]
	\includegraphics[scale=0.5]{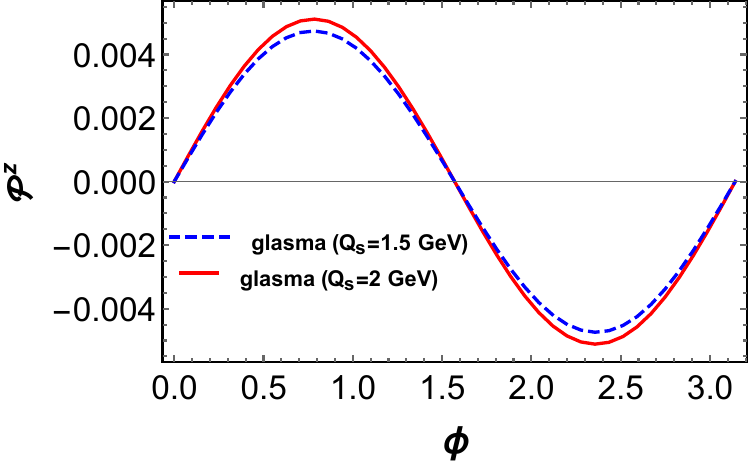}\\
	\vspace{10pt}
	\includegraphics[scale=0.5]{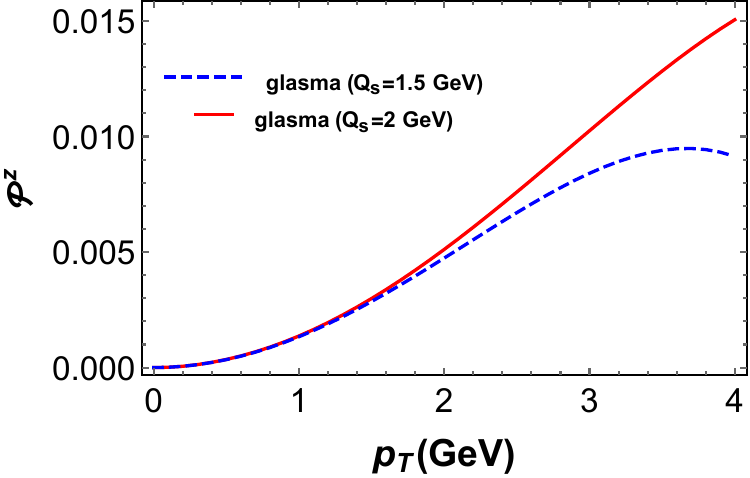}
	\caption{$\mathcal{P}^z$ at $y_p=0$ from glasma and Eq.~(\ref{eq:quark_Sch_dis}) for quark distributions. The upper panel shows $\phi$ dependence with $p_{T}=2$ GeV and the lower shows $p_{T}$ dependence with $\phi=\pi/4$. The blue dashed and red solid curves correspond to $Q_s=1.5$ GeV and $Q_s=2$ GeV, respectively.}\label{fig:GL_Sch_Pz}
\end{figure}
\begin{figure}[t]
	\includegraphics[scale=0.5]{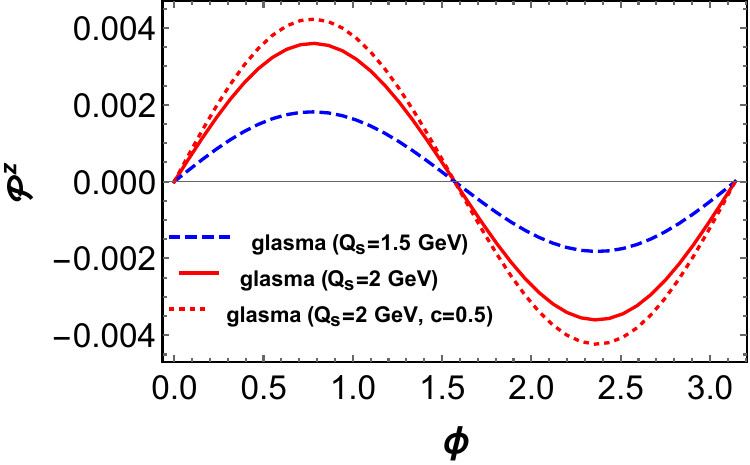}\\
    \vspace{10pt}
    \includegraphics[scale=0.5]{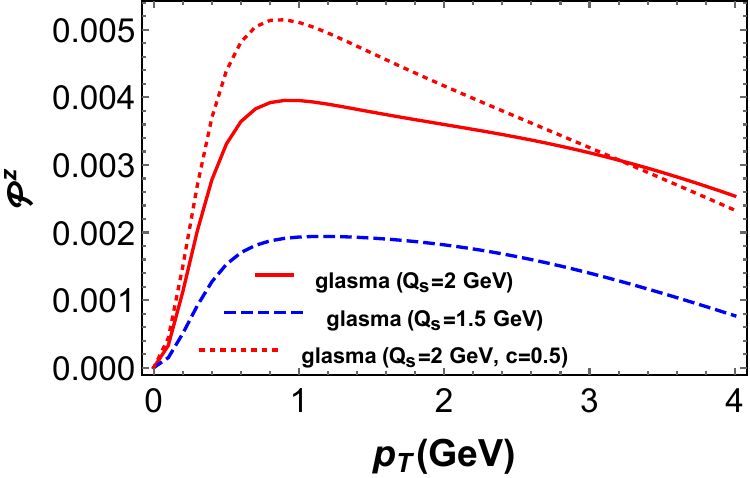}
	\caption{$\mathcal{P}^z$ at $y_p=0$ from glasma and Eq.~(\ref{eq:quark_dis}) for quark distributions. 
	The same color assignment as Fig.~\ref{fig:GL_Sch_Pz} for $c=1$.
	The additional red dotted curves represent the cases with $Q_s=2$ GeV and $c=0.5$ for comparisons.}\label{fig:GL_Pz}
\end{figure}
We consider $\epsilon=0$ and a weak initial flow with $u_{T}\ll 1$ such that $|{\bm p\cdot \bm u}|\ll Q_s$, whereby $f^{\rm s}_V$ and $N_{v}$ are almost $\Phi$ independent. In such a case, the contribution from $G_2$ in $\int d\Sigma\cdot pu^{[y}p^{x]}u^0$ becomes dominant, which results in
\begin{eqnarray}
\int d\Sigma\cdot pu^{[y}p^{x]}u^0
\approx u_{T}p_{T}^2\pi\sin 2\phi\int 
	\frac{dr d\eta r^3
	\tau\delta}{r_m^2}
\end{eqnarray}
and consequently the $\sin 2\phi$ structure is manifested. We note that the polarization vanishes for isotropic flow, $\delta=0$ as expected. 

We have numerically evaluated Eq.~(\ref{eq:Pz_glasma_origin}) by integrating over $r_{m}-\Delta t\leq r\leq r_{m}$, $0\leq\Phi\leq 2\pi$, and $-1\leq \eta\leq 1$ at a fixed $\tilde{\tau}_{f}$, where the regime $r_{m}-\Delta t\leq r\leq r_{m}$ is set to capture only the corona contribution. Here we focus on the case with $\tilde{\tau}_{f}\approx 0$ such that $\tau_{f}\approx r\leq r_{m}$. We take $\hbar=1$, $N_c=3$, $m=0.5$ GeV as the constituent strange quark mass, $M_{\Lambda}=1.12$ GeV, $\Delta t=0.2$ fm and $r_m\approx \tau^{\rm max}_{f}=0.5$ fm as the typical time scale smaller than $1$ fm for glasma phase. We set $\delta=0.3$ by analogy with the anisotropy parameter adopted in the thermal model for non-central collisions \cite{Florkowski:2019voj,Kumar:2023ojl}, while we take $u_{T}=0.01$ and $u_T=0.2$ for Eq.~(\ref{eq:quark_Sch_dis}) and Eq.~(\ref{eq:quark_dis}) characterizing the initial and final quark distributions in the glasma phase, respectively, as the weak non-equilibrium flow gradually generated by pressure gradients of glasma in time. The $\phi$ dependence of $\mathcal{P}^z$ at fixed $p_{T}=2$ GeV for $Q_s=1.5$ or $2$ GeV are shown in Fig.~\ref{fig:GL_Sch_Pz} and Fig.~\ref{fig:GL_Pz}, where the $\sin 2\phi$ structure is observed as anticipated. In addition, $\mathcal{P}^z$ with transverse-momentum dependence at fixed $\phi=\pi/4$ is also presented therein. For large $p_{T}$, the $G_1$ contribution becomes more substantial when $|{\bm p}\cdot{\bm u}|\sim Q_s$, which may cause the sign flipping of $\mathcal{P}^z$ at sufficiently large $p_{T}$. 

\sect{Polarization from quark gluon plasma}
One may similarly consider LSP engendered by isotropic color fields coming from the soft thermal gluons in QGP. For more realistic setup of the flow profile, we may employ the thermal model with single freeze-out for HIC \cite{Broniowski:2001we}. Following this approach, we set
\begin{eqnarray}
	u^{\mu}=\frac{1}{N_{u}}\big(t,x\sqrt{1+\delta},y\sqrt{1-\delta},z\big),
\end{eqnarray}  
where $N_{u}=\sqrt{\tilde{\tau}^2-(x^2-y^2)\delta }$ and $\delta$ again is a model parameter delineating the transverse anisotropy of the flow. For the isotropic color-field correlators, we approximate $\langle B^{a}_{z}(x)B^a_{z}(x)\rangle\approx\langle E^{a}_{\rm T}(x)E^a_{\rm T}(x)\rangle\approx T^4$ albeit more rigorous estimation needed. Then, we may evaluate $\mathcal{P}^z$ by simply multiplying Eq.~(\ref{eq:Pz_glasma_origin}) by $4g^2N_c/(N_c^2-1)$ and replacing $Q_s$ with $T$ and adopt the Fermi-Dirac distribution for quarks, while now integrating over the range $0 \leq r \leq r_{m}$.

\begin{figure}[t]
	\includegraphics[scale=0.5]{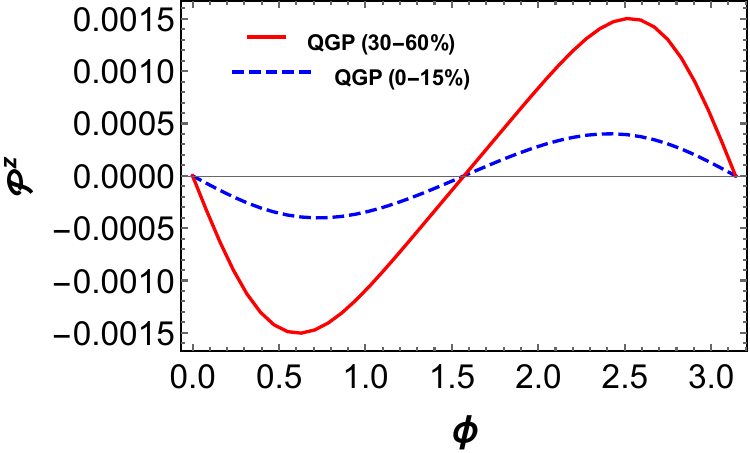}\quad\includegraphics[scale=0.5]{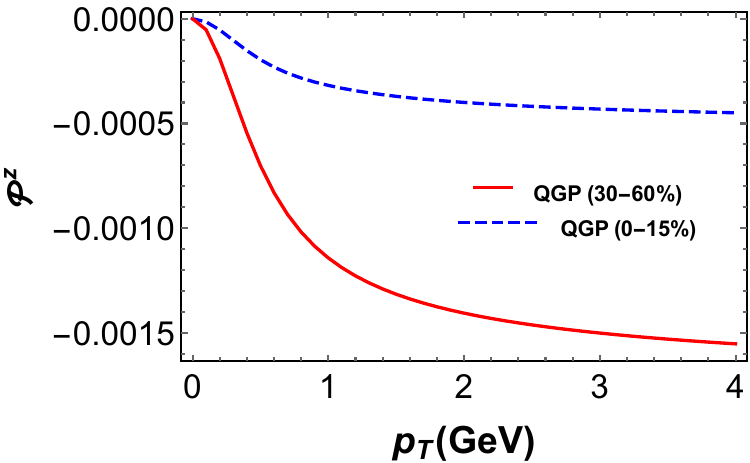}
	\caption{$\mathcal{P}^z$ at $y_p=0$ from QGP. The upper panel shows $\phi$ dependence with $p_{T}=2$ GeV and the lower shows $p_{T}$ dependence with $\phi=\pi/4$. The blue dashed and red solid curves correspond to $0-15$\% centrality and $30-60$\% centrality, respectively.}\label{fig:QGP_Pz}
\end{figure}

Following Refs.~\cite{Florkowski:2019voj,Kumar:2023ojl} for fitting the hadron spectra and flow of Au-Au collisions at $\sqrt{s_{NN}}=130$ GeV, we take $\epsilon=0.055$, $\delta=0.12$, $\tilde{\tau}_{f}=7.7$ fm, and $r_{m}=6.5$ fm for $0-15$\% centrality and $\epsilon=0.137$, $\delta=0.37$, $\tilde{\tau}_{f}=4.3$ fm, and $r_{m}=3.8$ fm for $30-60$\% centrality, respectively. We also adopt $T=T_{f}=165$ MeV as the freeze-out temperature, $g=\sqrt{4\pi/3}$, and $\Delta t=0.2$ fm on an equal footing to glasma (also such that $T_{f}$ is approximately constant). The $\phi$ and $p_{T}$ dependence for $\mathcal{P}^z$ are shown in Fig.~\ref{fig:QGP_Pz}. Since now $|{\bm p}\cdot {\bm u}|\gg T_{f}$, the $G_1$ contribution dominates and the overall sign of $\mathcal{P}^{z}$ is opposite with respect to the glasma case, while the $\sin 2\phi$ structure remains. However, the overall magnitude here could have larger uncertainties subject to chosen parameters. We only emphasize the qualitative difference with the glasma case.   

\textit{Discussions and summary}.---
Overall, the sign and azimuthal-angle dependence of $\mathcal{P}^z$ obtained from glasma fields are consistent with present experimental observations \cite{STAR:2019erd,CMS:2025nqr}. Although more sophisticated modeling is needed to quantify $\mathcal{P}^z$, our rough order-of-magnitude estimate for $\mathcal{P}^z\sim 0.1-1$\% is also comparable to the measurements. In practice, we expect that the observed polarization has contributions from both the glasma (corona) and QGP (core). Accordingly, the total degree of polarization may be expressed as
\begin{eqnarray}\label{eq:total_Pz}
\mathcal{P}^z=\frac{N_{\rm GL}\mathcal{P}^z_{\rm GL}+N_{\rm QGP}\mathcal{P}^z_{\rm QGP}}{N_{\rm GL}+N_{\rm QGP}},
\end{eqnarray}  
where $N$ denotes the particle number and the subscripts $\rm GL$ and $\rm QGP$ represent the contributions from the glasma (corona) and QGP (core), respectively. 
In general, $N_{\rm GL}$ and $N_{\rm QGP}$ should be more dominant at high $p_{T}$ for small systems and at low $p_{T}$ for large systems, respectively. More sophisticated modeling and simulations for the core and corona scenario similar to Ref.~\cite{Kanakubo:2021qcw} will be needed to quantify the results. Despite the dependence on transverse momenta and centrality, it is found therein that the corona contribution for hadron yields is about 50\% when the multiplicity per rapidity reaches $\mathcal{O}(10^1)$ for both p+p and Pb+Pb collisions. For a heuristic estimation, by taking $\mathcal{P}^z_{\rm GL}=4\times 10^{-3}$ and $\mathcal{P}^z_{\rm QGP}=-1.5\times 10^{-3}$ from Fig.~\ref{fig:GL_Sch_Pz} and Fig.~\ref{fig:QGP_Pz}, Eq.~(\ref{eq:total_Pz}) still yields $\mathcal{P}^z\approx 0.1\%$ for multiplicity around tens to a few hundreds as for small collision systems.

Recent measured strength of $\mathcal{P}^{z}$ in proton-nucleus collisions at LHC shows a mild opposite trend against multiplicity \cite{CMS:2025nqr}, which could be qualitatively explained by the substantial contribution from the corona in smaller collision systems due to the suppression of $N_{\rm QGP}$ with lower multiplicity. When the local equilibrium condition for spin degrees of freedom is reached, there exist additional contributions from thermal vorticity and thermal-shear corrections for $\mathcal{P}^{z}_{\rm QGP}$, further competing with the glasma effect. Finally, we remark that the initial flow for quarks in glasma could also stem from mini-jets and thus the momentum anisotropy and hydrodynamic flow may be misaligned with increasing multiplicity.    

In conclusion, we have shown that the correlators of color fields can create local spin polarization in the presence of anisotropic flow. We further argued that this mechanism can qualitatively explain the LSP of $\Lambda/\bar{\Lambda}$ hyperons in high-energy nuclear collisions from the glasma fields in corona, especially for small collision systems, while the polarization could be partially modified by the thermal color fields or hydrodynamic effects in QGP. Our findings demonstrate the possible significant influence of gluons on spin transport of quarks in HIC and complement the studies on spin alignment of vector mesons by strong forces \cite{Sheng:2022wsy,Sheng:2022ffb,Sheng:2023urn,Kumar:2022ylt,Kumar:2023ghs,Yang:2024qpy}. Extending previous applications of QCD effective kinetic theory  \cite{Arnold:2002zm,Kurkela:2015qoa,Kurkela:2018wud} to derive the initial conditions for hydrodynamics, our approach could provide the initial conditions for spin transport theories in the QGP such as spin hydrodynamics \cite{Florkowski:2018fap,Huang:2024ffg}.

\acknowledgments
\textit{Acknowledgments}. 
D.-L. Y. thanks A. Kumar for useful discussions.
This work was supported by a
grant from the U. S. Department of Energy, Office of Science (DE-FG02-05ER41367), by National Science and Technology Council (Taiwan) under Grant No. NSTC 113-2628-M-001-009-MY4, and by Academia Sinica under Project No. AS-CDA-114-M01.

\bibliography{local_polarization_color_fields_PRL_accepted.bbl}

\begin{thebibliography}{96}%
\makeatletter
\providecommand \@ifxundefined [1]{%
 \@ifx{#1\undefined}
}%
\providecommand \@ifnum [1]{%
 \ifnum #1\expandafter \@firstoftwo
 \else \expandafter \@secondoftwo
 \fi
}%
\providecommand \@ifx [1]{%
 \ifx #1\expandafter \@firstoftwo
 \else \expandafter \@secondoftwo
 \fi
}%
\providecommand \natexlab [1]{#1}%
\providecommand \enquote  [1]{``#1''}%
\providecommand \bibnamefont  [1]{#1}%
\providecommand \bibfnamefont [1]{#1}%
\providecommand \citenamefont [1]{#1}%
\providecommand \href@noop [0]{\@secondoftwo}%
\providecommand \href [0]{\begingroup \@sanitize@url \@href}%
\providecommand \@href[1]{\@@startlink{#1}\@@href}%
\providecommand \@@href[1]{\endgroup#1\@@endlink}%
\providecommand \@sanitize@url [0]{\catcode `\\12\catcode `\$12\catcode
  `\&12\catcode `\#12\catcode `\^12\catcode `\_12\catcode `\%12\relax}%
\providecommand \@@startlink[1]{}%
\providecommand \@@endlink[0]{}%
\providecommand \url  [0]{\begingroup\@sanitize@url \@url }%
\providecommand \@url [1]{\endgroup\@href {#1}{\urlprefix }}%
\providecommand \urlprefix  [0]{URL }%
\providecommand \Eprint [0]{\href }%
\providecommand \doibase [0]{https://doi.org/}%
\providecommand \selectlanguage [0]{\@gobble}%
\providecommand \bibinfo  [0]{\@secondoftwo}%
\providecommand \bibfield  [0]{\@secondoftwo}%
\providecommand \translation [1]{[#1]}%
\providecommand \BibitemOpen [0]{}%
\providecommand \bibitemStop [0]{}%
\providecommand \bibitemNoStop [0]{.\EOS\space}%
\providecommand \EOS [0]{\spacefactor3000\relax}%
\providecommand \BibitemShut  [1]{\csname bibitem#1\endcsname}%
\let\auto@bib@innerbib\@empty
\bibitem [{\citenamefont {Adamczyk}\ \emph {et~al.}(2017)\citenamefont
  {Adamczyk} \emph {et~al.}}]{STAR:2017ckg}%
  \BibitemOpen
  \bibfield  {author} {\bibinfo {author} {\bibfnamefont {L.}~\bibnamefont
  {Adamczyk}} \emph {et~al.} (\bibinfo {collaboration} {STAR}),\ }\bibfield
  {title} {\bibinfo {title} {{Global $\Lambda$ hyperon polarization in nuclear
  collisions: evidence for the most vortical fluid}},\ }\href
  {https://doi.org/10.1038/nature23004} {\bibfield  {journal} {\bibinfo
  {journal} {Nature}\ }\textbf {\bibinfo {volume} {548}},\ \bibinfo {pages}
  {62} (\bibinfo {year} {2017})},\ \Eprint {https://arxiv.org/abs/1701.06657}
  {arXiv:1701.06657 [nucl-ex]} \BibitemShut {NoStop}%
\bibitem [{\citenamefont {Adam}\ \emph {et~al.}(2018)\citenamefont {Adam} \emph
  {et~al.}}]{STAR:2018gyt}%
  \BibitemOpen
  \bibfield  {author} {\bibinfo {author} {\bibfnamefont {J.}~\bibnamefont
  {Adam}} \emph {et~al.} (\bibinfo {collaboration} {STAR}),\ }\bibfield
  {title} {\bibinfo {title} {{Global polarization of $\Lambda$ hyperons in
  Au+Au collisions at $\sqrt{s_{_{NN}}}$ = 200 GeV}},\ }\href
  {https://doi.org/10.1103/PhysRevC.98.014910} {\bibfield  {journal} {\bibinfo
  {journal} {Phys. Rev. C}\ }\textbf {\bibinfo {volume} {98}},\ \bibinfo
  {pages} {014910} (\bibinfo {year} {2018})},\ \Eprint
  {https://arxiv.org/abs/1805.04400} {arXiv:1805.04400 [nucl-ex]} \BibitemShut
  {NoStop}%
\bibitem [{\citenamefont {Liang}\ and\ \citenamefont
  {Wang}(2005{\natexlab{a}})}]{Liang:2004ph}%
  \BibitemOpen
  \bibfield  {author} {\bibinfo {author} {\bibfnamefont {Z.-T.}\ \bibnamefont
  {Liang}}\ and\ \bibinfo {author} {\bibfnamefont {X.-N.}\ \bibnamefont
  {Wang}},\ }\bibfield  {title} {\bibinfo {title} {{Globally polarized
  quark-gluon plasma in non-central A+A collisions}},\ }\href
  {https://doi.org/10.1103/PhysRevLett.94.102301,
  10.1103/PhysRevLett.96.039901} {\bibfield  {journal} {\bibinfo  {journal}
  {Phys. Rev. Lett.}\ }\textbf {\bibinfo {volume} {94}},\ \bibinfo {pages}
  {102301} (\bibinfo {year} {2005}{\natexlab{a}})},\ \bibinfo {note} {[Erratum:
  Phys. Rev. Lett.96,039901(2006)]},\ \Eprint
  {https://arxiv.org/abs/nucl-th/0410079} {arXiv:nucl-th/0410079 [nucl-th]}
  \BibitemShut {NoStop}%
\bibitem [{\citenamefont {Liang}\ and\ \citenamefont
  {Wang}(2005{\natexlab{b}})}]{Liang:2004xn}%
  \BibitemOpen
  \bibfield  {author} {\bibinfo {author} {\bibfnamefont {Z.-T.}\ \bibnamefont
  {Liang}}\ and\ \bibinfo {author} {\bibfnamefont {X.-N.}\ \bibnamefont
  {Wang}},\ }\bibfield  {title} {\bibinfo {title} {{Spin alignment of vector
  mesons in non-central A+A collisions}},\ }\href
  {https://doi.org/10.1016/j.physletb.2005.09.060} {\bibfield  {journal}
  {\bibinfo  {journal} {Phys. Lett. B}\ }\textbf {\bibinfo {volume} {629}},\
  \bibinfo {pages} {20} (\bibinfo {year} {2005}{\natexlab{b}})},\ \Eprint
  {https://arxiv.org/abs/nucl-th/0411101} {arXiv:nucl-th/0411101} \BibitemShut
  {NoStop}%
\bibitem [{\citenamefont {Becattini}\ \emph
  {et~al.}(2013{\natexlab{a}})\citenamefont {Becattini}, \citenamefont
  {Chandra}, \citenamefont {Del~Zanna},\ and\ \citenamefont
  {Grossi}}]{Becattini2013a}%
  \BibitemOpen
  \bibfield  {author} {\bibinfo {author} {\bibfnamefont {F.}~\bibnamefont
  {Becattini}}, \bibinfo {author} {\bibfnamefont {V.}~\bibnamefont {Chandra}},
  \bibinfo {author} {\bibfnamefont {L.}~\bibnamefont {Del~Zanna}},\ and\
  \bibinfo {author} {\bibfnamefont {E.}~\bibnamefont {Grossi}},\ }\bibfield
  {title} {\bibinfo {title} {{Relativistic distribution function for particles
  with spin at local thermodynamical equilibrium}},\ }\href
  {https://doi.org/10.1016/j.aop.2013.07.004} {\bibfield  {journal} {\bibinfo
  {journal} {Annals Phys.}\ }\textbf {\bibinfo {volume} {338}},\ \bibinfo
  {pages} {32} (\bibinfo {year} {2013}{\natexlab{a}})},\ \Eprint
  {https://arxiv.org/abs/1303.3431} {arXiv:1303.3431 [nucl-th]} \BibitemShut
  {NoStop}%
\bibitem [{\citenamefont {Fang}\ \emph {et~al.}(2016)\citenamefont {Fang},
  \citenamefont {Pang}, \citenamefont {Wang},\ and\ \citenamefont
  {Wang}}]{Fang:2016vpj}%
  \BibitemOpen
  \bibfield  {author} {\bibinfo {author} {\bibfnamefont {R.-h.}\ \bibnamefont
  {Fang}}, \bibinfo {author} {\bibfnamefont {L.-g.}\ \bibnamefont {Pang}},
  \bibinfo {author} {\bibfnamefont {Q.}~\bibnamefont {Wang}},\ and\ \bibinfo
  {author} {\bibfnamefont {X.-n.}\ \bibnamefont {Wang}},\ }\bibfield  {title}
  {\bibinfo {title} {{Polarization of massive fermions in a vortical fluid}},\
  }\href {https://doi.org/10.1103/PhysRevC.94.024904} {\bibfield  {journal}
  {\bibinfo  {journal} {Phys. Rev. C}\ }\textbf {\bibinfo {volume} {94}},\
  \bibinfo {pages} {024904} (\bibinfo {year} {2016})},\ \Eprint
  {https://arxiv.org/abs/1604.04036} {arXiv:1604.04036 [nucl-th]} \BibitemShut
  {NoStop}%
\bibitem [{\citenamefont {Becattini}\ \emph
  {et~al.}(2013{\natexlab{b}})\citenamefont {Becattini}, \citenamefont
  {Csernai},\ and\ \citenamefont {Wang}}]{Becattini:2013vja}%
  \BibitemOpen
  \bibfield  {author} {\bibinfo {author} {\bibfnamefont {F.}~\bibnamefont
  {Becattini}}, \bibinfo {author} {\bibfnamefont {L.}~\bibnamefont {Csernai}},\
  and\ \bibinfo {author} {\bibfnamefont {D.~J.}\ \bibnamefont {Wang}},\
  }\bibfield  {title} {\bibinfo {title} {{$\Lambda$ polarization in peripheral
  heavy ion collisions}},\ }\href {https://doi.org/10.1103/PhysRevC.93.069901,
  10.1103/PhysRevC.88.034905} {\bibfield  {journal} {\bibinfo  {journal} {Phys.
  Rev.}\ }\textbf {\bibinfo {volume} {C88}},\ \bibinfo {pages} {034905}
  (\bibinfo {year} {2013}{\natexlab{b}})},\ \bibinfo {note} {[Erratum: Phys.
  Rev. C93,no.6,069901(2016)]},\ \Eprint {https://arxiv.org/abs/1304.4427}
  {arXiv:1304.4427 [nucl-th]} \BibitemShut {NoStop}%
\bibitem [{\citenamefont {Hidaka}\ \emph {et~al.}(2018)\citenamefont {Hidaka},
  \citenamefont {Pu},\ and\ \citenamefont {Yang}}]{Hidaka:2017auj}%
  \BibitemOpen
  \bibfield  {author} {\bibinfo {author} {\bibfnamefont {Y.}~\bibnamefont
  {Hidaka}}, \bibinfo {author} {\bibfnamefont {S.}~\bibnamefont {Pu}},\ and\
  \bibinfo {author} {\bibfnamefont {D.-L.}\ \bibnamefont {Yang}},\ }\bibfield
  {title} {\bibinfo {title} {{Nonlinear Responses of Chiral Fluids from Kinetic
  Theory}},\ }\href {https://doi.org/10.1103/PhysRevD.97.016004} {\bibfield
  {journal} {\bibinfo  {journal} {Phys. Rev.}\ }\textbf {\bibinfo {volume}
  {D97}},\ \bibinfo {pages} {016004} (\bibinfo {year} {2018})},\ \Eprint
  {https://arxiv.org/abs/1710.00278} {arXiv:1710.00278 [hep-th]} \BibitemShut
  {NoStop}%
\bibitem [{\citenamefont {Liu}\ and\ \citenamefont
  {Yin}(2021{\natexlab{a}})}]{Liu:2020dxg}%
  \BibitemOpen
  \bibfield  {author} {\bibinfo {author} {\bibfnamefont {S.~Y.~F.}\
  \bibnamefont {Liu}}\ and\ \bibinfo {author} {\bibfnamefont {Y.}~\bibnamefont
  {Yin}},\ }\bibfield  {title} {\bibinfo {title} {{Spin Hall effect in
  heavy-ion collisions}},\ }\href {https://doi.org/10.1103/PhysRevD.104.054043}
  {\bibfield  {journal} {\bibinfo  {journal} {Phys. Rev. D}\ }\textbf {\bibinfo
  {volume} {104}},\ \bibinfo {pages} {054043} (\bibinfo {year}
  {2021}{\natexlab{a}})},\ \Eprint {https://arxiv.org/abs/2006.12421}
  {arXiv:2006.12421 [nucl-th]} \BibitemShut {NoStop}%
\bibitem [{\citenamefont {Liu}\ and\ \citenamefont
  {Yin}(2021{\natexlab{b}})}]{Liu:2021uhn}%
  \BibitemOpen
  \bibfield  {author} {\bibinfo {author} {\bibfnamefont {S.~Y.~F.}\
  \bibnamefont {Liu}}\ and\ \bibinfo {author} {\bibfnamefont {Y.}~\bibnamefont
  {Yin}},\ }\bibfield  {title} {\bibinfo {title} {{Spin polarization induced by
  the hydrodynamic gradients}},\ }\href
  {https://doi.org/10.1007/JHEP07(2021)188} {\bibfield  {journal} {\bibinfo
  {journal} {JHEP}\ }\textbf {\bibinfo {volume} {07}},\ \bibinfo {pages}
  {188}},\ \Eprint {https://arxiv.org/abs/2103.09200} {arXiv:2103.09200
  [hep-ph]} \BibitemShut {NoStop}%
\bibitem [{\citenamefont {Becattini}\ \emph
  {et~al.}(2021{\natexlab{a}})\citenamefont {Becattini}, \citenamefont
  {Buzzegoli},\ and\ \citenamefont {Palermo}}]{Becattini:2021suc}%
  \BibitemOpen
  \bibfield  {author} {\bibinfo {author} {\bibfnamefont {F.}~\bibnamefont
  {Becattini}}, \bibinfo {author} {\bibfnamefont {M.}~\bibnamefont
  {Buzzegoli}},\ and\ \bibinfo {author} {\bibfnamefont {A.}~\bibnamefont
  {Palermo}},\ }\bibfield  {title} {\bibinfo {title} {{Spin-thermal shear
  coupling in a relativistic fluid}},\ }\href
  {https://doi.org/10.1016/j.physletb.2021.136519} {\bibfield  {journal}
  {\bibinfo  {journal} {Phys. Lett. B}\ }\textbf {\bibinfo {volume} {820}},\
  \bibinfo {pages} {136519} (\bibinfo {year} {2021}{\natexlab{a}})},\ \Eprint
  {https://arxiv.org/abs/2103.10917} {arXiv:2103.10917 [nucl-th]} \BibitemShut
  {NoStop}%
\bibitem [{\citenamefont {Fu}\ \emph {et~al.}(2021)\citenamefont {Fu},
  \citenamefont {Liu}, \citenamefont {Pang}, \citenamefont {Song},\ and\
  \citenamefont {Yin}}]{Fu:2021pok}%
  \BibitemOpen
  \bibfield  {author} {\bibinfo {author} {\bibfnamefont {B.}~\bibnamefont
  {Fu}}, \bibinfo {author} {\bibfnamefont {S.~Y.~F.}\ \bibnamefont {Liu}},
  \bibinfo {author} {\bibfnamefont {L.}~\bibnamefont {Pang}}, \bibinfo {author}
  {\bibfnamefont {H.}~\bibnamefont {Song}},\ and\ \bibinfo {author}
  {\bibfnamefont {Y.}~\bibnamefont {Yin}},\ }\bibfield  {title} {\bibinfo
  {title} {{Shear-Induced Spin Polarization in Heavy-Ion Collisions}},\ }\href
  {https://doi.org/10.1103/PhysRevLett.127.142301} {\bibfield  {journal}
  {\bibinfo  {journal} {Phys. Rev. Lett.}\ }\textbf {\bibinfo {volume} {127}},\
  \bibinfo {pages} {142301} (\bibinfo {year} {2021})},\ \Eprint
  {https://arxiv.org/abs/2103.10403} {arXiv:2103.10403 [hep-ph]} \BibitemShut
  {NoStop}%
\bibitem [{\citenamefont {Becattini}\ \emph
  {et~al.}(2021{\natexlab{b}})\citenamefont {Becattini}, \citenamefont
  {Buzzegoli}, \citenamefont {Inghirami}, \citenamefont {Karpenko},\ and\
  \citenamefont {Palermo}}]{Becattini:2021iol}%
  \BibitemOpen
  \bibfield  {author} {\bibinfo {author} {\bibfnamefont {F.}~\bibnamefont
  {Becattini}}, \bibinfo {author} {\bibfnamefont {M.}~\bibnamefont
  {Buzzegoli}}, \bibinfo {author} {\bibfnamefont {G.}~\bibnamefont
  {Inghirami}}, \bibinfo {author} {\bibfnamefont {I.}~\bibnamefont
  {Karpenko}},\ and\ \bibinfo {author} {\bibfnamefont {A.}~\bibnamefont
  {Palermo}},\ }\bibfield  {title} {\bibinfo {title} {{Local Polarization and
  Isothermal Local Equilibrium in Relativistic Heavy Ion Collisions}},\ }\href
  {https://doi.org/10.1103/PhysRevLett.127.272302} {\bibfield  {journal}
  {\bibinfo  {journal} {Phys. Rev. Lett.}\ }\textbf {\bibinfo {volume} {127}},\
  \bibinfo {pages} {272302} (\bibinfo {year} {2021}{\natexlab{b}})},\ \Eprint
  {https://arxiv.org/abs/2103.14621} {arXiv:2103.14621 [nucl-th]} \BibitemShut
  {NoStop}%
\bibitem [{\citenamefont {Yi}\ \emph {et~al.}(2021)\citenamefont {Yi},
  \citenamefont {Pu},\ and\ \citenamefont {Yang}}]{Yi:2021ryh}%
  \BibitemOpen
  \bibfield  {author} {\bibinfo {author} {\bibfnamefont {C.}~\bibnamefont
  {Yi}}, \bibinfo {author} {\bibfnamefont {S.}~\bibnamefont {Pu}},\ and\
  \bibinfo {author} {\bibfnamefont {D.-L.}\ \bibnamefont {Yang}},\ }\bibfield
  {title} {\bibinfo {title} {{Reexamination of local spin polarization beyond
  global equilibrium in relativistic heavy ion collisions}},\ }\href
  {https://doi.org/10.1103/PhysRevC.104.064901} {\bibfield  {journal} {\bibinfo
   {journal} {Phys. Rev. C}\ }\textbf {\bibinfo {volume} {104}},\ \bibinfo
  {pages} {064901} (\bibinfo {year} {2021})},\ \Eprint
  {https://arxiv.org/abs/2106.00238} {arXiv:2106.00238 [hep-ph]} \BibitemShut
  {NoStop}%
\bibitem [{\citenamefont {Ryu}\ \emph {et~al.}(2021)\citenamefont {Ryu},
  \citenamefont {Jupic},\ and\ \citenamefont {Shen}}]{Ryu:2021lnx}%
  \BibitemOpen
  \bibfield  {author} {\bibinfo {author} {\bibfnamefont {S.}~\bibnamefont
  {Ryu}}, \bibinfo {author} {\bibfnamefont {V.}~\bibnamefont {Jupic}},\ and\
  \bibinfo {author} {\bibfnamefont {C.}~\bibnamefont {Shen}},\ }\bibfield
  {title} {\bibinfo {title} {{Probing early-time longitudinal dynamics with the
  \ensuremath{\Lambda} hyperon's spin polarization in relativistic heavy-ion
  collisions}},\ }\href {https://doi.org/10.1103/PhysRevC.104.054908}
  {\bibfield  {journal} {\bibinfo  {journal} {Phys. Rev. C}\ }\textbf {\bibinfo
  {volume} {104}},\ \bibinfo {pages} {054908} (\bibinfo {year} {2021})},\
  \Eprint {https://arxiv.org/abs/2106.08125} {arXiv:2106.08125 [nucl-th]}
  \BibitemShut {NoStop}%
\bibitem [{\citenamefont {Florkowski}\ \emph {et~al.}(2022)\citenamefont
  {Florkowski}, \citenamefont {Kumar}, \citenamefont {Mazeliauskas},\ and\
  \citenamefont {Ryblewski}}]{Florkowski:2021xvy}%
  \BibitemOpen
  \bibfield  {author} {\bibinfo {author} {\bibfnamefont {W.}~\bibnamefont
  {Florkowski}}, \bibinfo {author} {\bibfnamefont {A.}~\bibnamefont {Kumar}},
  \bibinfo {author} {\bibfnamefont {A.}~\bibnamefont {Mazeliauskas}},\ and\
  \bibinfo {author} {\bibfnamefont {R.}~\bibnamefont {Ryblewski}},\ }\bibfield
  {title} {\bibinfo {title} {{Effect of thermal shear on longitudinal spin
  polarization in a thermal model}},\ }\href
  {https://doi.org/10.1103/PhysRevC.105.064901} {\bibfield  {journal} {\bibinfo
   {journal} {Phys. Rev. C}\ }\textbf {\bibinfo {volume} {105}},\ \bibinfo
  {pages} {064901} (\bibinfo {year} {2022})},\ \Eprint
  {https://arxiv.org/abs/2112.02799} {arXiv:2112.02799 [hep-ph]} \BibitemShut
  {NoStop}%
\bibitem [{\citenamefont {Sahoo}\ \emph {et~al.}(2025)\citenamefont {Sahoo},
  \citenamefont {Singh},\ and\ \citenamefont {Sahoo}}]{Sahoo:2024egx}%
  \BibitemOpen
  \bibfield  {author} {\bibinfo {author} {\bibfnamefont {B.}~\bibnamefont
  {Sahoo}}, \bibinfo {author} {\bibfnamefont {C.~R.}\ \bibnamefont {Singh}},\
  and\ \bibinfo {author} {\bibfnamefont {R.}~\bibnamefont {Sahoo}},\ }\bibfield
   {title} {\bibinfo {title} {{Estimating longitudinal polarization of
  $\Lambda$ and $\bar{\Lambda}$ hyperons at relativistic energies using
  hydrodynamic and transport models}},\ }\href
  {https://doi.org/10.1088/1402-4896/add9ef} {\bibfield  {journal} {\bibinfo
  {journal} {Phys. Scripta}\ }\textbf {\bibinfo {volume} {100}},\ \bibinfo
  {pages} {065310} (\bibinfo {year} {2025})},\ \Eprint
  {https://arxiv.org/abs/2404.15138} {arXiv:2404.15138 [hep-ph]} \BibitemShut
  {NoStop}%
\bibitem [{\citenamefont {Adam}\ \emph {et~al.}(2019)\citenamefont {Adam} \emph
  {et~al.}}]{STAR:2019erd}%
  \BibitemOpen
  \bibfield  {author} {\bibinfo {author} {\bibfnamefont {J.}~\bibnamefont
  {Adam}} \emph {et~al.} (\bibinfo {collaboration} {STAR}),\ }\bibfield
  {title} {\bibinfo {title} {{Polarization of $\Lambda$ ($\bar{\Lambda}$)
  hyperons along the beam direction in Au+Au collisions at $\sqrt{s_{_{NN}}}$ =
  200 GeV}},\ }\href {https://doi.org/10.1103/PhysRevLett.123.132301}
  {\bibfield  {journal} {\bibinfo  {journal} {Phys. Rev. Lett.}\ }\textbf
  {\bibinfo {volume} {123}},\ \bibinfo {pages} {132301} (\bibinfo {year}
  {2019})},\ \Eprint {https://arxiv.org/abs/1905.11917} {arXiv:1905.11917
  [nucl-ex]} \BibitemShut {NoStop}%
\bibitem [{\citenamefont {Liu}\ \emph {et~al.}(2020)\citenamefont {Liu},
  \citenamefont {Sun},\ and\ \citenamefont {Ko}}]{Liu:2019krs}%
  \BibitemOpen
  \bibfield  {author} {\bibinfo {author} {\bibfnamefont {S.~Y.}\ \bibnamefont
  {Liu}}, \bibinfo {author} {\bibfnamefont {Y.}~\bibnamefont {Sun}},\ and\
  \bibinfo {author} {\bibfnamefont {C.~M.}\ \bibnamefont {Ko}},\ }\bibfield
  {title} {\bibinfo {title} {{Spin Polarizations in a Covariant
  Angular-Momentum-Conserved Chiral Transport Model}},\ }\href
  {https://doi.org/10.1103/PhysRevLett.125.062301} {\bibfield  {journal}
  {\bibinfo  {journal} {Phys. Rev. Lett.}\ }\textbf {\bibinfo {volume} {125}},\
  \bibinfo {pages} {062301} (\bibinfo {year} {2020})},\ \Eprint
  {https://arxiv.org/abs/1910.06774} {arXiv:1910.06774 [nucl-th]} \BibitemShut
  {NoStop}%
\bibitem [{\citenamefont {Sun}\ and\ \citenamefont {Yan}(2025)}]{Sun:2024isb}%
  \BibitemOpen
  \bibfield  {author} {\bibinfo {author} {\bibfnamefont {J.-A.}\ \bibnamefont
  {Sun}}\ and\ \bibinfo {author} {\bibfnamefont {L.}~\bibnamefont {Yan}},\
  }\bibfield  {title} {\bibinfo {title} {{Weak magnetic effect in quark-gluon
  plasma and local spin polarization*}},\ }\href
  {https://doi.org/10.1088/1674-1137/adc120} {\bibfield  {journal} {\bibinfo
  {journal} {Chin. Phys. C}\ }\textbf {\bibinfo {volume} {49}},\ \bibinfo
  {pages} {071001} (\bibinfo {year} {2025})},\ \Eprint
  {https://arxiv.org/abs/2401.07458} {arXiv:2401.07458 [nucl-th]} \BibitemShut
  {NoStop}%
\bibitem [{\citenamefont {Yi}\ \emph {et~al.}(2025)\citenamefont {Yi},
  \citenamefont {Wu}, \citenamefont {Zhu}, \citenamefont {Pu},\ and\
  \citenamefont {Qin}}]{Yi:2024kwu}%
  \BibitemOpen
  \bibfield  {author} {\bibinfo {author} {\bibfnamefont {C.}~\bibnamefont
  {Yi}}, \bibinfo {author} {\bibfnamefont {X.-Y.}\ \bibnamefont {Wu}}, \bibinfo
  {author} {\bibfnamefont {J.}~\bibnamefont {Zhu}}, \bibinfo {author}
  {\bibfnamefont {S.}~\bibnamefont {Pu}},\ and\ \bibinfo {author}
  {\bibfnamefont {G.-Y.}\ \bibnamefont {Qin}},\ }\bibfield  {title} {\bibinfo
  {title} {{Spin polarization of {\ensuremath{\Lambda}} hyperons along the beam
  direction in p+Pb collisions at sNN=8.16 TeV using hydrodynamic
  approaches}},\ }\href {https://doi.org/10.1103/PhysRevC.111.044901}
  {\bibfield  {journal} {\bibinfo  {journal} {Phys. Rev. C}\ }\textbf {\bibinfo
  {volume} {111}},\ \bibinfo {pages} {044901} (\bibinfo {year} {2025})},\
  \Eprint {https://arxiv.org/abs/2408.04296} {arXiv:2408.04296 [hep-ph]}
  \BibitemShut {NoStop}%
\bibitem [{\citenamefont {Hayrapetyan}\ \emph {et~al.}(2025)\citenamefont
  {Hayrapetyan} \emph {et~al.}}]{CMS:2025nqr}%
  \BibitemOpen
  \bibfield  {author} {\bibinfo {author} {\bibfnamefont {A.}~\bibnamefont
  {Hayrapetyan}} \emph {et~al.} (\bibinfo {collaboration} {CMS}),\ }\bibfield
  {title} {\bibinfo {title} {{Observation of $\Lambda$ hyperon local
  polarization in pPb collisions at $\sqrt{s_\mathrm{NN}}$ = 8.16 TeV}},\
  }\href {https://doi.org/10.1103/6ywq-gm61} {\bibfield  {journal} {\bibinfo
  {journal} {Phys. Rev. Lett.}\ }\textbf {\bibinfo {volume} {135}},\ \bibinfo
  {pages} {132301} (\bibinfo {year} {2025})},\ \Eprint
  {https://arxiv.org/abs/2502.07898} {arXiv:2502.07898 [nucl-ex]} \BibitemShut
  {NoStop}%
\bibitem [{\citenamefont {Lin}\ and\ \citenamefont {Wang}(2022)}]{Lin:2022tma}%
  \BibitemOpen
  \bibfield  {author} {\bibinfo {author} {\bibfnamefont {S.}~\bibnamefont
  {Lin}}\ and\ \bibinfo {author} {\bibfnamefont {Z.}~\bibnamefont {Wang}},\
  }\bibfield  {title} {\bibinfo {title} {{Shear induced polarization:
  collisional contributions}},\ }\href
  {https://doi.org/10.1007/JHEP12(2022)030} {\bibfield  {journal} {\bibinfo
  {journal} {JHEP}\ }\textbf {\bibinfo {volume} {12}},\ \bibinfo {pages}
  {030}},\ \Eprint {https://arxiv.org/abs/2206.12573} {arXiv:2206.12573
  [hep-ph]} \BibitemShut {NoStop}%
\bibitem [{\citenamefont {Fang}\ \emph {et~al.}(2022)\citenamefont {Fang},
  \citenamefont {Pu},\ and\ \citenamefont {Yang}}]{Fang:2022ttm}%
  \BibitemOpen
  \bibfield  {author} {\bibinfo {author} {\bibfnamefont {S.}~\bibnamefont
  {Fang}}, \bibinfo {author} {\bibfnamefont {S.}~\bibnamefont {Pu}},\ and\
  \bibinfo {author} {\bibfnamefont {D.-L.}\ \bibnamefont {Yang}},\ }\bibfield
  {title} {\bibinfo {title} {{Quantum kinetic theory for dynamical spin
  polarization from QED-type interaction}},\ }\href
  {https://doi.org/10.1103/PhysRevD.106.016002} {\bibfield  {journal} {\bibinfo
   {journal} {Phys. Rev. D}\ }\textbf {\bibinfo {volume} {106}},\ \bibinfo
  {pages} {016002} (\bibinfo {year} {2022})},\ \Eprint
  {https://arxiv.org/abs/2204.11519} {arXiv:2204.11519 [hep-ph]} \BibitemShut
  {NoStop}%
\bibitem [{\citenamefont {Fang}\ \emph {et~al.}(2024)\citenamefont {Fang},
  \citenamefont {Pu},\ and\ \citenamefont {Yang}}]{Fang:2023bbw}%
  \BibitemOpen
  \bibfield  {author} {\bibinfo {author} {\bibfnamefont {S.}~\bibnamefont
  {Fang}}, \bibinfo {author} {\bibfnamefont {S.}~\bibnamefont {Pu}},\ and\
  \bibinfo {author} {\bibfnamefont {D.-L.}\ \bibnamefont {Yang}},\ }\bibfield
  {title} {\bibinfo {title} {{Spin polarization and spin alignment from quantum
  kinetic theory with self-energy corrections}},\ }\href
  {https://doi.org/10.1103/PhysRevD.109.034034} {\bibfield  {journal} {\bibinfo
   {journal} {Phys. Rev. D}\ }\textbf {\bibinfo {volume} {109}},\ \bibinfo
  {pages} {034034} (\bibinfo {year} {2024})},\ \Eprint
  {https://arxiv.org/abs/2311.15197} {arXiv:2311.15197 [hep-ph]} \BibitemShut
  {NoStop}%
\bibitem [{\citenamefont {Lin}\ and\ \citenamefont {Wang}(2025)}]{Lin:2024zik}%
  \BibitemOpen
  \bibfield  {author} {\bibinfo {author} {\bibfnamefont {S.}~\bibnamefont
  {Lin}}\ and\ \bibinfo {author} {\bibfnamefont {Z.}~\bibnamefont {Wang}},\
  }\bibfield  {title} {\bibinfo {title} {Steady state, displacement current,
  and spin polarization for massless fermion in a shear flow},\ }\href
  {https://doi.org/10.1103/PhysRevD.111.034032} {\bibfield  {journal} {\bibinfo
   {journal} {Phys. Rev. D}\ }\textbf {\bibinfo {volume} {111}},\ \bibinfo
  {pages} {034032} (\bibinfo {year} {2025})},\ \Eprint
  {https://arxiv.org/abs/2406.10003} {arXiv:2406.10003 [hep-ph]} \BibitemShut
  {NoStop}%
\bibitem [{\citenamefont {Fang}\ and\ \citenamefont {Pu}(2025)}]{Fang:2024vds}%
  \BibitemOpen
  \bibfield  {author} {\bibinfo {author} {\bibfnamefont {S.}~\bibnamefont
  {Fang}}\ and\ \bibinfo {author} {\bibfnamefont {S.}~\bibnamefont {Pu}},\
  }\bibfield  {title} {\bibinfo {title} {{Collisional corrections to spin
  polarization from quantum kinetic theory using Chapman-Enskog expansion}},\
  }\href {https://doi.org/10.1103/PhysRevD.111.034015} {\bibfield  {journal}
  {\bibinfo  {journal} {Phys. Rev. D}\ }\textbf {\bibinfo {volume} {111}},\
  \bibinfo {pages} {034015} (\bibinfo {year} {2025})},\ \Eprint
  {https://arxiv.org/abs/2408.09877} {arXiv:2408.09877 [hep-ph]} \BibitemShut
  {NoStop}%
\bibitem [{\citenamefont {Lin}\ and\ \citenamefont {Tian}(2024)}]{Lin:2024svh}%
  \BibitemOpen
  \bibfield  {author} {\bibinfo {author} {\bibfnamefont {S.}~\bibnamefont
  {Lin}}\ and\ \bibinfo {author} {\bibfnamefont {J.}~\bibnamefont {Tian}},\
  }\href@noop {} {\bibinfo {title} {{Spin Polarized Quasi-particle in
  Off-equilibrium Medium}}} (\bibinfo {year} {2024}),\ \Eprint
  {https://arxiv.org/abs/2410.22935} {arXiv:2410.22935 [hep-ph]} \BibitemShut
  {NoStop}%
\bibitem [{\citenamefont {Wang}\ and\ \citenamefont
  {Lin}(2025)}]{Wang:2024lis}%
  \BibitemOpen
  \bibfield  {author} {\bibinfo {author} {\bibfnamefont {Z.}~\bibnamefont
  {Wang}}\ and\ \bibinfo {author} {\bibfnamefont {S.}~\bibnamefont {Lin}},\
  }\bibfield  {title} {\bibinfo {title} {{Spin polarization for massive fermion
  in a shear flow: Complete results at O({\ensuremath{\partial}})}},\ }\href
  {https://doi.org/10.1103/PhysRevD.111.056023} {\bibfield  {journal} {\bibinfo
   {journal} {Phys. Rev. D}\ }\textbf {\bibinfo {volume} {111}},\ \bibinfo
  {pages} {056023} (\bibinfo {year} {2025})},\ \Eprint
  {https://arxiv.org/abs/2411.19550} {arXiv:2411.19550 [hep-ph]} \BibitemShut
  {NoStop}%
\bibitem [{\citenamefont {Fang}\ \emph {et~al.}(2025)\citenamefont {Fang},
  \citenamefont {Pu},\ and\ \citenamefont {Yang}}]{Fang:2025pzy}%
  \BibitemOpen
  \bibfield  {author} {\bibinfo {author} {\bibfnamefont {S.}~\bibnamefont
  {Fang}}, \bibinfo {author} {\bibfnamefont {S.}~\bibnamefont {Pu}},\ and\
  \bibinfo {author} {\bibfnamefont {D.-L.}\ \bibnamefont {Yang}},\ }\href@noop
  {} {\bibinfo {title} {{Radiative corrections on vortical spin polarization in
  hot QCD matter}}} (\bibinfo {year} {2025}),\ \Eprint
  {https://arxiv.org/abs/2503.13320} {arXiv:2503.13320 [hep-ph]} \BibitemShut
  {NoStop}%
\bibitem [{\citenamefont {Weickgenannt}\ \emph
  {et~al.}(2022{\natexlab{a}})\citenamefont {Weickgenannt}, \citenamefont
  {Wagner}, \citenamefont {Speranza},\ and\ \citenamefont
  {Rischke}}]{Weickgenannt:2022zxs}%
  \BibitemOpen
  \bibfield  {author} {\bibinfo {author} {\bibfnamefont {N.}~\bibnamefont
  {Weickgenannt}}, \bibinfo {author} {\bibfnamefont {D.}~\bibnamefont
  {Wagner}}, \bibinfo {author} {\bibfnamefont {E.}~\bibnamefont {Speranza}},\
  and\ \bibinfo {author} {\bibfnamefont {D.~H.}\ \bibnamefont {Rischke}},\
  }\bibfield  {title} {\bibinfo {title} {{Relativistic second-order dissipative
  spin hydrodynamics from the method of moments}},\ }\href
  {https://doi.org/10.1103/PhysRevD.106.096014} {\bibfield  {journal} {\bibinfo
   {journal} {Phys. Rev. D}\ }\textbf {\bibinfo {volume} {106}},\ \bibinfo
  {pages} {096014} (\bibinfo {year} {2022}{\natexlab{a}})},\ \Eprint
  {https://arxiv.org/abs/2203.04766} {arXiv:2203.04766 [nucl-th]} \BibitemShut
  {NoStop}%
\bibitem [{\citenamefont {Weickgenannt}\ \emph
  {et~al.}(2022{\natexlab{b}})\citenamefont {Weickgenannt}, \citenamefont
  {Wagner}, \citenamefont {Speranza},\ and\ \citenamefont
  {Rischke}}]{Weickgenannt:2022qvh}%
  \BibitemOpen
  \bibfield  {author} {\bibinfo {author} {\bibfnamefont {N.}~\bibnamefont
  {Weickgenannt}}, \bibinfo {author} {\bibfnamefont {D.}~\bibnamefont
  {Wagner}}, \bibinfo {author} {\bibfnamefont {E.}~\bibnamefont {Speranza}},\
  and\ \bibinfo {author} {\bibfnamefont {D.~H.}\ \bibnamefont {Rischke}},\
  }\bibfield  {title} {\bibinfo {title} {{Relativistic dissipative spin
  hydrodynamics from kinetic theory with a nonlocal collision term}},\ }\href
  {https://doi.org/10.1103/PhysRevD.106.L091901} {\bibfield  {journal}
  {\bibinfo  {journal} {Phys. Rev. D}\ }\textbf {\bibinfo {volume} {106}},\
  \bibinfo {pages} {L091901} (\bibinfo {year} {2022}{\natexlab{b}})},\ \Eprint
  {https://arxiv.org/abs/2208.01955} {arXiv:2208.01955 [nucl-th]} \BibitemShut
  {NoStop}%
\bibitem [{\citenamefont {Weickgenannt}\ and\ \citenamefont
  {Blaizot}(2025)}]{Weickgenannt:2024ibf}%
  \BibitemOpen
  \bibfield  {author} {\bibinfo {author} {\bibfnamefont {N.}~\bibnamefont
  {Weickgenannt}}\ and\ \bibinfo {author} {\bibfnamefont {J.-P.}\ \bibnamefont
  {Blaizot}},\ }\bibfield  {title} {\bibinfo {title} {{Spin kinetic theory with
  a nonlocal relaxation time approximation}},\ }\href
  {https://doi.org/10.1103/PhysRevD.111.056006} {\bibfield  {journal} {\bibinfo
   {journal} {Phys. Rev. D}\ }\textbf {\bibinfo {volume} {111}},\ \bibinfo
  {pages} {056006} (\bibinfo {year} {2025})},\ \Eprint
  {https://arxiv.org/abs/2409.11045} {arXiv:2409.11045 [hep-ph]} \BibitemShut
  {NoStop}%
\bibitem [{\citenamefont {Wagner}(2025)}]{Wagner:2024fry}%
  \BibitemOpen
  \bibfield  {author} {\bibinfo {author} {\bibfnamefont {D.}~\bibnamefont
  {Wagner}},\ }\bibfield  {title} {\bibinfo {title} {{Resummed spin
  hydrodynamics from quantum kinetic theory}},\ }\href
  {https://doi.org/10.1103/PhysRevD.111.016008} {\bibfield  {journal} {\bibinfo
   {journal} {Phys. Rev. D}\ }\textbf {\bibinfo {volume} {111}},\ \bibinfo
  {pages} {016008} (\bibinfo {year} {2025})},\ \Eprint
  {https://arxiv.org/abs/2409.07143} {arXiv:2409.07143 [nucl-th]} \BibitemShut
  {NoStop}%
\bibitem [{\citenamefont {Sapna}\ \emph {et~al.}(2025)\citenamefont {Sapna},
  \citenamefont {Singh},\ and\ \citenamefont {Wagner}}]{Sapna:2025yss}%
  \BibitemOpen
  \bibfield  {author} {\bibinfo {author} {\bibnamefont {Sapna}}, \bibinfo
  {author} {\bibfnamefont {S.~K.}\ \bibnamefont {Singh}},\ and\ \bibinfo
  {author} {\bibfnamefont {D.}~\bibnamefont {Wagner}},\ }\href@noop {}
  {\bibinfo {title} {{Spin Polarization of $\Lambda$ hyperons from Dissipative
  Spin Hydrodynamics}}} (\bibinfo {year} {2025}),\ \Eprint
  {https://arxiv.org/abs/2503.22552} {arXiv:2503.22552 [hep-ph]} \BibitemShut
  {NoStop}%
\bibitem [{\citenamefont {Stephanov}\ and\ \citenamefont
  {Yin}(2012)}]{Stephanov:2012ki}%
  \BibitemOpen
  \bibfield  {author} {\bibinfo {author} {\bibfnamefont {M.}~\bibnamefont
  {Stephanov}}\ and\ \bibinfo {author} {\bibfnamefont {Y.}~\bibnamefont
  {Yin}},\ }\bibfield  {title} {\bibinfo {title} {{Chiral Kinetic Theory}},\
  }\href {https://doi.org/10.1103/PhysRevLett.109.162001} {\bibfield  {journal}
  {\bibinfo  {journal} {Phys. Rev. Lett.}\ }\textbf {\bibinfo {volume} {109}},\
  \bibinfo {pages} {162001} (\bibinfo {year} {2012})},\ \Eprint
  {https://arxiv.org/abs/1207.0747} {arXiv:1207.0747 [hep-th]} \BibitemShut
  {NoStop}%
\bibitem [{\citenamefont {Son}\ and\ \citenamefont
  {Yamamoto}(2012)}]{Son:2012wh}%
  \BibitemOpen
  \bibfield  {author} {\bibinfo {author} {\bibfnamefont {D.~T.}\ \bibnamefont
  {Son}}\ and\ \bibinfo {author} {\bibfnamefont {N.}~\bibnamefont {Yamamoto}},\
  }\bibfield  {title} {\bibinfo {title} {{Berry Curvature, Triangle Anomalies,
  and the Chiral Magnetic Effect in Fermi Liquids}},\ }\href
  {https://doi.org/10.1103/PhysRevLett.109.181602} {\bibfield  {journal}
  {\bibinfo  {journal} {Phys. Rev. Lett.}\ }\textbf {\bibinfo {volume} {109}},\
  \bibinfo {pages} {181602} (\bibinfo {year} {2012})},\ \Eprint
  {https://arxiv.org/abs/1203.2697} {arXiv:1203.2697 [cond-mat.mes-hall]}
  \BibitemShut {NoStop}%
\bibitem [{\citenamefont {Chen}\ \emph {et~al.}(2013)\citenamefont {Chen},
  \citenamefont {Pu}, \citenamefont {Wang},\ and\ \citenamefont
  {Wang}}]{Chen:2012ca}%
  \BibitemOpen
  \bibfield  {author} {\bibinfo {author} {\bibfnamefont {J.-W.}\ \bibnamefont
  {Chen}}, \bibinfo {author} {\bibfnamefont {S.}~\bibnamefont {Pu}}, \bibinfo
  {author} {\bibfnamefont {Q.}~\bibnamefont {Wang}},\ and\ \bibinfo {author}
  {\bibfnamefont {X.-N.}\ \bibnamefont {Wang}},\ }\bibfield  {title} {\bibinfo
  {title} {{Berry Curvature and Four-Dimensional Monopoles in the Relativistic
  Chiral Kinetic Equation}},\ }\href
  {https://doi.org/10.1103/PhysRevLett.110.262301} {\bibfield  {journal}
  {\bibinfo  {journal} {Phys. Rev. Lett.}\ }\textbf {\bibinfo {volume} {110}},\
  \bibinfo {pages} {262301} (\bibinfo {year} {2013})},\ \Eprint
  {https://arxiv.org/abs/1210.8312} {arXiv:1210.8312 [hep-th]} \BibitemShut
  {NoStop}%
\bibitem [{\citenamefont {Hidaka}\ \emph {et~al.}(2017)\citenamefont {Hidaka},
  \citenamefont {Pu},\ and\ \citenamefont {Yang}}]{Hidaka:2016yjf}%
  \BibitemOpen
  \bibfield  {author} {\bibinfo {author} {\bibfnamefont {Y.}~\bibnamefont
  {Hidaka}}, \bibinfo {author} {\bibfnamefont {S.}~\bibnamefont {Pu}},\ and\
  \bibinfo {author} {\bibfnamefont {D.-L.}\ \bibnamefont {Yang}},\ }\bibfield
  {title} {\bibinfo {title} {{Relativistic Chiral Kinetic Theory from Quantum
  Field Theories}},\ }\href {https://doi.org/10.1103/PhysRevD.95.091901}
  {\bibfield  {journal} {\bibinfo  {journal} {Phys. Rev.}\ }\textbf {\bibinfo
  {volume} {D95}},\ \bibinfo {pages} {091901} (\bibinfo {year} {2017})},\
  \Eprint {https://arxiv.org/abs/1612.04630} {arXiv:1612.04630 [hep-th]}
  \BibitemShut {NoStop}%
\bibitem [{\citenamefont {Gao}\ and\ \citenamefont
  {Liang}(2019)}]{Gao:2019znl}%
  \BibitemOpen
  \bibfield  {author} {\bibinfo {author} {\bibfnamefont {J.-H.}\ \bibnamefont
  {Gao}}\ and\ \bibinfo {author} {\bibfnamefont {Z.-T.}\ \bibnamefont
  {Liang}},\ }\bibfield  {title} {\bibinfo {title} {{Relativistic Quantum
  Kinetic Theory for Massive Fermions and Spin Effects}},\ }\href
  {https://doi.org/10.1103/PhysRevD.100.056021} {\bibfield  {journal} {\bibinfo
   {journal} {Phys. Rev.}\ }\textbf {\bibinfo {volume} {D100}},\ \bibinfo
  {pages} {056021} (\bibinfo {year} {2019})},\ \Eprint
  {https://arxiv.org/abs/1902.06510} {arXiv:1902.06510 [hep-ph]} \BibitemShut
  {NoStop}%
\bibitem [{\citenamefont {Weickgenannt}\ \emph {et~al.}(2019)\citenamefont
  {Weickgenannt}, \citenamefont {Sheng}, \citenamefont {Speranza},
  \citenamefont {Wang},\ and\ \citenamefont {Rischke}}]{Weickgenannt:2019dks}%
  \BibitemOpen
  \bibfield  {author} {\bibinfo {author} {\bibfnamefont {N.}~\bibnamefont
  {Weickgenannt}}, \bibinfo {author} {\bibfnamefont {X.-L.}\ \bibnamefont
  {Sheng}}, \bibinfo {author} {\bibfnamefont {E.}~\bibnamefont {Speranza}},
  \bibinfo {author} {\bibfnamefont {Q.}~\bibnamefont {Wang}},\ and\ \bibinfo
  {author} {\bibfnamefont {D.~H.}\ \bibnamefont {Rischke}},\ }\bibfield
  {title} {\bibinfo {title} {{Kinetic theory for massive spin-1/2 particles
  from the Wigner-function formalism}},\ }\href
  {https://doi.org/10.1103/PhysRevD.100.056018} {\bibfield  {journal} {\bibinfo
   {journal} {Phys. Rev.}\ }\textbf {\bibinfo {volume} {D100}},\ \bibinfo
  {pages} {056018} (\bibinfo {year} {2019})},\ \Eprint
  {https://arxiv.org/abs/1902.06513} {arXiv:1902.06513 [hep-ph]} \BibitemShut
  {NoStop}%
\bibitem [{\citenamefont {Hattori}\ \emph
  {et~al.}(2019{\natexlab{a}})\citenamefont {Hattori}, \citenamefont {Hidaka},\
  and\ \citenamefont {Yang}}]{Hattori:2019ahi}%
  \BibitemOpen
  \bibfield  {author} {\bibinfo {author} {\bibfnamefont {K.}~\bibnamefont
  {Hattori}}, \bibinfo {author} {\bibfnamefont {Y.}~\bibnamefont {Hidaka}},\
  and\ \bibinfo {author} {\bibfnamefont {D.-L.}\ \bibnamefont {Yang}},\
  }\bibfield  {title} {\bibinfo {title} {{Axial Kinetic Theory and Spin
  Transport for Fermions with Arbitrary Mass}},\ }\href
  {https://doi.org/10.1103/PhysRevD.100.096011} {\bibfield  {journal} {\bibinfo
   {journal} {Phys. Rev.}\ }\textbf {\bibinfo {volume} {D100}},\ \bibinfo
  {pages} {096011} (\bibinfo {year} {2019}{\natexlab{a}})},\ \Eprint
  {https://arxiv.org/abs/1903.01653} {arXiv:1903.01653 [hep-ph]} \BibitemShut
  {NoStop}%
\bibitem [{\citenamefont {Wang}\ \emph {et~al.}(2019)\citenamefont {Wang},
  \citenamefont {Guo}, \citenamefont {Shi},\ and\ \citenamefont
  {Zhuang}}]{Wang:2019moi}%
  \BibitemOpen
  \bibfield  {author} {\bibinfo {author} {\bibfnamefont {Z.}~\bibnamefont
  {Wang}}, \bibinfo {author} {\bibfnamefont {X.}~\bibnamefont {Guo}}, \bibinfo
  {author} {\bibfnamefont {S.}~\bibnamefont {Shi}},\ and\ \bibinfo {author}
  {\bibfnamefont {P.}~\bibnamefont {Zhuang}},\ }\bibfield  {title} {\bibinfo
  {title} {{Mass Correction to Chiral Kinetic Equations}},\ }\href
  {https://doi.org/10.1103/PhysRevD.100.014015} {\bibfield  {journal} {\bibinfo
   {journal} {Phys. Rev.}\ }\textbf {\bibinfo {volume} {D100}},\ \bibinfo
  {pages} {014015} (\bibinfo {year} {2019})},\ \Eprint
  {https://arxiv.org/abs/1903.03461} {arXiv:1903.03461 [hep-ph]} \BibitemShut
  {NoStop}%
\bibitem [{\citenamefont {Li}\ and\ \citenamefont {Yee}(2019)}]{Li:2019qkf}%
  \BibitemOpen
  \bibfield  {author} {\bibinfo {author} {\bibfnamefont {S.}~\bibnamefont
  {Li}}\ and\ \bibinfo {author} {\bibfnamefont {H.-U.}\ \bibnamefont {Yee}},\
  }\bibfield  {title} {\bibinfo {title} {{Quantum Kinetic Theory of Spin
  Polarization of Massive Quarks in Perturbative QCD: Leading Log}},\ }\href
  {https://doi.org/10.1103/PhysRevD.100.056022} {\bibfield  {journal} {\bibinfo
   {journal} {Phys. Rev.}\ }\textbf {\bibinfo {volume} {D100}},\ \bibinfo
  {pages} {056022} (\bibinfo {year} {2019})},\ \Eprint
  {https://arxiv.org/abs/1905.10463} {arXiv:1905.10463 [hep-ph]} \BibitemShut
  {NoStop}%
\bibitem [{\citenamefont {Yang}\ \emph {et~al.}(2020)\citenamefont {Yang},
  \citenamefont {Hattori},\ and\ \citenamefont {Hidaka}}]{Yang:2020hri}%
  \BibitemOpen
  \bibfield  {author} {\bibinfo {author} {\bibfnamefont {D.-L.}\ \bibnamefont
  {Yang}}, \bibinfo {author} {\bibfnamefont {K.}~\bibnamefont {Hattori}},\ and\
  \bibinfo {author} {\bibfnamefont {Y.}~\bibnamefont {Hidaka}},\ }\bibfield
  {title} {\bibinfo {title} {{Effective quantum kinetic theory for spin
  transport of fermions with collsional effects}},\ }\href
  {https://doi.org/10.1007/JHEP07(2020)070} {\bibfield  {journal} {\bibinfo
  {journal} {JHEP}\ }\textbf {\bibinfo {volume} {20}},\ \bibinfo {pages}
  {070}},\ \Eprint {https://arxiv.org/abs/2002.02612} {arXiv:2002.02612
  [hep-ph]} \BibitemShut {NoStop}%
\bibitem [{\citenamefont {Weickgenannt}\ \emph {et~al.}(2021)\citenamefont
  {Weickgenannt}, \citenamefont {Speranza}, \citenamefont {Sheng},
  \citenamefont {Wang},\ and\ \citenamefont {Rischke}}]{Weickgenannt:2020aaf}%
  \BibitemOpen
  \bibfield  {author} {\bibinfo {author} {\bibfnamefont {N.}~\bibnamefont
  {Weickgenannt}}, \bibinfo {author} {\bibfnamefont {E.}~\bibnamefont
  {Speranza}}, \bibinfo {author} {\bibfnamefont {X.-l.}\ \bibnamefont {Sheng}},
  \bibinfo {author} {\bibfnamefont {Q.}~\bibnamefont {Wang}},\ and\ \bibinfo
  {author} {\bibfnamefont {D.~H.}\ \bibnamefont {Rischke}},\ }\bibfield
  {title} {\bibinfo {title} {{Generating Spin Polarization from Vorticity
  through Nonlocal Collisions}},\ }\href
  {https://doi.org/10.1103/PhysRevLett.127.052301} {\bibfield  {journal}
  {\bibinfo  {journal} {Phys. Rev. Lett.}\ }\textbf {\bibinfo {volume} {127}},\
  \bibinfo {pages} {052301} (\bibinfo {year} {2021})},\ \Eprint
  {https://arxiv.org/abs/2005.01506} {arXiv:2005.01506 [hep-ph]} \BibitemShut
  {NoStop}%
\bibitem [{\citenamefont {Wang}\ \emph {et~al.}(2021)\citenamefont {Wang},
  \citenamefont {Guo},\ and\ \citenamefont {Zhuang}}]{Wang:2020pej}%
  \BibitemOpen
  \bibfield  {author} {\bibinfo {author} {\bibfnamefont {Z.}~\bibnamefont
  {Wang}}, \bibinfo {author} {\bibfnamefont {X.}~\bibnamefont {Guo}},\ and\
  \bibinfo {author} {\bibfnamefont {P.}~\bibnamefont {Zhuang}},\ }\bibfield
  {title} {\bibinfo {title} {{Equilibrium Spin Distribution From Detailed
  Balance}},\ }\href {https://doi.org/10.1140/epjc/s10052-021-09586-8}
  {\bibfield  {journal} {\bibinfo  {journal} {Eur. Phys. J. C}\ }\textbf
  {\bibinfo {volume} {81}},\ \bibinfo {pages} {799} (\bibinfo {year} {2021})},\
  \Eprint {https://arxiv.org/abs/2009.10930} {arXiv:2009.10930 [hep-th]}
  \BibitemShut {NoStop}%
\bibitem [{\citenamefont {Hattori}\ \emph {et~al.}(2021)\citenamefont
  {Hattori}, \citenamefont {Hidaka}, \citenamefont {Yamamoto},\ and\
  \citenamefont {Yang}}]{Hattori:2020gqh}%
  \BibitemOpen
  \bibfield  {author} {\bibinfo {author} {\bibfnamefont {K.}~\bibnamefont
  {Hattori}}, \bibinfo {author} {\bibfnamefont {Y.}~\bibnamefont {Hidaka}},
  \bibinfo {author} {\bibfnamefont {N.}~\bibnamefont {Yamamoto}},\ and\
  \bibinfo {author} {\bibfnamefont {D.-L.}\ \bibnamefont {Yang}},\ }\bibfield
  {title} {\bibinfo {title} {{Wigner functions and quantum kinetic theory of
  polarized photons}},\ }\href {https://doi.org/10.1007/JHEP02(2021)001}
  {\bibfield  {journal} {\bibinfo  {journal} {JHEP}\ }\textbf {\bibinfo
  {volume} {02}},\ \bibinfo {pages} {001}},\ \Eprint
  {https://arxiv.org/abs/2010.13368} {arXiv:2010.13368 [hep-ph]} \BibitemShut
  {NoStop}%
\bibitem [{\citenamefont {Hidaka}\ \emph {et~al.}(2022)\citenamefont {Hidaka},
  \citenamefont {Pu}, \citenamefont {Wang},\ and\ \citenamefont
  {Yang}}]{Hidaka:2022dmn}%
  \BibitemOpen
  \bibfield  {author} {\bibinfo {author} {\bibfnamefont {Y.}~\bibnamefont
  {Hidaka}}, \bibinfo {author} {\bibfnamefont {S.}~\bibnamefont {Pu}}, \bibinfo
  {author} {\bibfnamefont {Q.}~\bibnamefont {Wang}},\ and\ \bibinfo {author}
  {\bibfnamefont {D.-L.}\ \bibnamefont {Yang}},\ }\bibfield  {title} {\bibinfo
  {title} {{Foundations and applications of quantum kinetic theory}},\ }\href
  {https://doi.org/10.1016/j.ppnp.2022.103989} {\bibfield  {journal} {\bibinfo
  {journal} {Prog. Part. Nucl. Phys.}\ }\textbf {\bibinfo {volume} {127}},\
  \bibinfo {pages} {103989} (\bibinfo {year} {2022})},\ \Eprint
  {https://arxiv.org/abs/2201.07644} {arXiv:2201.07644 [hep-ph]} \BibitemShut
  {NoStop}%
\bibitem [{\citenamefont {Montenegro}\ \emph {et~al.}(2017)\citenamefont
  {Montenegro}, \citenamefont {Tinti},\ and\ \citenamefont
  {Torrieri}}]{Montenegro:2017rbu}%
  \BibitemOpen
  \bibfield  {author} {\bibinfo {author} {\bibfnamefont {D.}~\bibnamefont
  {Montenegro}}, \bibinfo {author} {\bibfnamefont {L.}~\bibnamefont {Tinti}},\
  and\ \bibinfo {author} {\bibfnamefont {G.}~\bibnamefont {Torrieri}},\
  }\bibfield  {title} {\bibinfo {title} {{Ideal relativistic fluid limit for a
  medium with polarization}},\ }\href
  {https://doi.org/10.1103/PhysRevD.96.079901, 10.1103/PhysRevD.96.056012}
  {\bibfield  {journal} {\bibinfo  {journal} {Phys. Rev.}\ }\textbf {\bibinfo
  {volume} {D96}},\ \bibinfo {pages} {056012} (\bibinfo {year} {2017})},\
  \bibinfo {note} {[Addendum: Phys. Rev.D96,no.7,079901(2017)]},\ \Eprint
  {https://arxiv.org/abs/1701.08263} {arXiv:1701.08263 [hep-th]} \BibitemShut
  {NoStop}%
\bibitem [{\citenamefont {Florkowski}\ \emph {et~al.}(2018)\citenamefont
  {Florkowski}, \citenamefont {Friman}, \citenamefont {Jaiswal},\ and\
  \citenamefont {Speranza}}]{Florkowski:2017ruc}%
  \BibitemOpen
  \bibfield  {author} {\bibinfo {author} {\bibfnamefont {W.}~\bibnamefont
  {Florkowski}}, \bibinfo {author} {\bibfnamefont {B.}~\bibnamefont {Friman}},
  \bibinfo {author} {\bibfnamefont {A.}~\bibnamefont {Jaiswal}},\ and\ \bibinfo
  {author} {\bibfnamefont {E.}~\bibnamefont {Speranza}},\ }\bibfield  {title}
  {\bibinfo {title} {{Relativistic fluid dynamics with spin}},\ }\href
  {https://doi.org/10.1103/PhysRevC.97.041901} {\bibfield  {journal} {\bibinfo
  {journal} {Phys. Rev.}\ }\textbf {\bibinfo {volume} {C97}},\ \bibinfo {pages}
  {041901} (\bibinfo {year} {2018})},\ \Eprint
  {https://arxiv.org/abs/1705.00587} {arXiv:1705.00587 [nucl-th]} \BibitemShut
  {NoStop}%
\bibitem [{\citenamefont {Florkowski}\ \emph
  {et~al.}(2019{\natexlab{a}})\citenamefont {Florkowski}, \citenamefont
  {Ryblewski},\ and\ \citenamefont {Kumar}}]{Florkowski:2018fap}%
  \BibitemOpen
  \bibfield  {author} {\bibinfo {author} {\bibfnamefont {W.}~\bibnamefont
  {Florkowski}}, \bibinfo {author} {\bibfnamefont {R.}~\bibnamefont
  {Ryblewski}},\ and\ \bibinfo {author} {\bibfnamefont {A.}~\bibnamefont
  {Kumar}},\ }\bibfield  {title} {\bibinfo {title} {{Relativistic hydrodynamics
  for spin-polarized fluids}},\ }\href
  {https://doi.org/10.1016/j.ppnp.2019.07.001} {\bibfield  {journal} {\bibinfo
  {journal} {Prog. Part. Nucl. Phys.}\ }\textbf {\bibinfo {volume} {108}},\
  \bibinfo {pages} {103709} (\bibinfo {year} {2019}{\natexlab{a}})},\ \Eprint
  {https://arxiv.org/abs/1811.04409} {arXiv:1811.04409 [nucl-th]} \BibitemShut
  {NoStop}%
\bibitem [{\citenamefont {Yang}(2018)}]{Yang:2018lew}%
  \BibitemOpen
  \bibfield  {author} {\bibinfo {author} {\bibfnamefont {D.-L.}\ \bibnamefont
  {Yang}},\ }\bibfield  {title} {\bibinfo {title} {{Side-Jump Induced
  Spin-Orbit Interaction of Chiral Fluids from Kinetic Theory}},\ }\href
  {https://doi.org/10.1103/PhysRevD.98.076019} {\bibfield  {journal} {\bibinfo
  {journal} {Phys. Rev.}\ }\textbf {\bibinfo {volume} {D98}},\ \bibinfo {pages}
  {076019} (\bibinfo {year} {2018})},\ \Eprint
  {https://arxiv.org/abs/1807.02395} {arXiv:1807.02395 [nucl-th]} \BibitemShut
  {NoStop}%
\bibitem [{\citenamefont {Hattori}\ \emph
  {et~al.}(2019{\natexlab{b}})\citenamefont {Hattori}, \citenamefont {Hongo},
  \citenamefont {Huang}, \citenamefont {Matsuo},\ and\ \citenamefont
  {Taya}}]{Hattori:2019lfp}%
  \BibitemOpen
  \bibfield  {author} {\bibinfo {author} {\bibfnamefont {K.}~\bibnamefont
  {Hattori}}, \bibinfo {author} {\bibfnamefont {M.}~\bibnamefont {Hongo}},
  \bibinfo {author} {\bibfnamefont {X.-G.}\ \bibnamefont {Huang}}, \bibinfo
  {author} {\bibfnamefont {M.}~\bibnamefont {Matsuo}},\ and\ \bibinfo {author}
  {\bibfnamefont {H.}~\bibnamefont {Taya}},\ }\bibfield  {title} {\bibinfo
  {title} {{Fate of spin polarization in a relativistic fluid: An
  entropy-current analysis}},\ }\href
  {https://doi.org/10.1016/j.physletb.2019.05.040} {\bibfield  {journal}
  {\bibinfo  {journal} {Phys. Lett.}\ }\textbf {\bibinfo {volume} {B795}},\
  \bibinfo {pages} {100} (\bibinfo {year} {2019}{\natexlab{b}})},\ \Eprint
  {https://arxiv.org/abs/1901.06615} {arXiv:1901.06615 [hep-th]} \BibitemShut
  {NoStop}%
\bibitem [{\citenamefont {Fukushima}\ and\ \citenamefont
  {Pu}(2021)}]{Fukushima:2020ucl}%
  \BibitemOpen
  \bibfield  {author} {\bibinfo {author} {\bibfnamefont {K.}~\bibnamefont
  {Fukushima}}\ and\ \bibinfo {author} {\bibfnamefont {S.}~\bibnamefont {Pu}},\
  }\bibfield  {title} {\bibinfo {title} {{Spin hydrodynamics and symmetric
  energy-momentum tensors \textendash{} A current induced by the spin vorticity
  \textendash{}}},\ }\href {https://doi.org/10.1016/j.physletb.2021.136346}
  {\bibfield  {journal} {\bibinfo  {journal} {Phys. Lett. B}\ }\textbf
  {\bibinfo {volume} {817}},\ \bibinfo {pages} {136346} (\bibinfo {year}
  {2021})},\ \Eprint {https://arxiv.org/abs/2010.01608} {arXiv:2010.01608
  [hep-th]} \BibitemShut {NoStop}%
\bibitem [{\citenamefont {Shi}\ \emph {et~al.}(2021)\citenamefont {Shi},
  \citenamefont {Gale},\ and\ \citenamefont {Jeon}}]{Shi:2020htn}%
  \BibitemOpen
  \bibfield  {author} {\bibinfo {author} {\bibfnamefont {S.}~\bibnamefont
  {Shi}}, \bibinfo {author} {\bibfnamefont {C.}~\bibnamefont {Gale}},\ and\
  \bibinfo {author} {\bibfnamefont {S.}~\bibnamefont {Jeon}},\ }\bibfield
  {title} {\bibinfo {title} {{From chiral kinetic theory to relativistic
  viscous spin hydrodynamics}},\ }\href
  {https://doi.org/10.1103/PhysRevC.103.044906} {\bibfield  {journal} {\bibinfo
   {journal} {Phys. Rev. C}\ }\textbf {\bibinfo {volume} {103}},\ \bibinfo
  {pages} {044906} (\bibinfo {year} {2021})},\ \Eprint
  {https://arxiv.org/abs/2008.08618} {arXiv:2008.08618 [nucl-th]} \BibitemShut
  {NoStop}%
\bibitem [{\citenamefont {Li}\ \emph {et~al.}(2021)\citenamefont {Li},
  \citenamefont {Stephanov},\ and\ \citenamefont {Yee}}]{Li:2020eon}%
  \BibitemOpen
  \bibfield  {author} {\bibinfo {author} {\bibfnamefont {S.}~\bibnamefont
  {Li}}, \bibinfo {author} {\bibfnamefont {M.~A.}\ \bibnamefont {Stephanov}},\
  and\ \bibinfo {author} {\bibfnamefont {H.-U.}\ \bibnamefont {Yee}},\
  }\bibfield  {title} {\bibinfo {title} {{Nondissipative Second-Order
  Transport, Spin, and Pseudogauge Transformations in Hydrodynamics}},\ }\href
  {https://doi.org/10.1103/PhysRevLett.127.082302} {\bibfield  {journal}
  {\bibinfo  {journal} {Phys. Rev. Lett.}\ }\textbf {\bibinfo {volume} {127}},\
  \bibinfo {pages} {082302} (\bibinfo {year} {2021})},\ \Eprint
  {https://arxiv.org/abs/2011.12318} {arXiv:2011.12318 [hep-th]} \BibitemShut
  {NoStop}%
\bibitem [{\citenamefont {Hongo}\ \emph {et~al.}(2021)\citenamefont {Hongo},
  \citenamefont {Huang}, \citenamefont {Kaminski}, \citenamefont {Stephanov},\
  and\ \citenamefont {Yee}}]{Hongo:2021ona}%
  \BibitemOpen
  \bibfield  {author} {\bibinfo {author} {\bibfnamefont {M.}~\bibnamefont
  {Hongo}}, \bibinfo {author} {\bibfnamefont {X.-G.}\ \bibnamefont {Huang}},
  \bibinfo {author} {\bibfnamefont {M.}~\bibnamefont {Kaminski}}, \bibinfo
  {author} {\bibfnamefont {M.}~\bibnamefont {Stephanov}},\ and\ \bibinfo
  {author} {\bibfnamefont {H.-U.}\ \bibnamefont {Yee}},\ }\bibfield  {title}
  {\bibinfo {title} {{Relativistic spin hydrodynamics with torsion and linear
  response theory for spin relaxation}},\ }\href
  {https://doi.org/10.1007/JHEP11(2021)150} {\bibfield  {journal} {\bibinfo
  {journal} {JHEP}\ }\textbf {\bibinfo {volume} {11}},\ \bibinfo {pages}
  {150}},\ \Eprint {https://arxiv.org/abs/2107.14231} {arXiv:2107.14231
  [hep-th]} \BibitemShut {NoStop}%
\bibitem [{\citenamefont {Bhadury}\ \emph {et~al.}(2022)\citenamefont
  {Bhadury}, \citenamefont {Florkowski}, \citenamefont {Jaiswal}, \citenamefont
  {Kumar},\ and\ \citenamefont {Ryblewski}}]{Bhadury:2022ulr}%
  \BibitemOpen
  \bibfield  {author} {\bibinfo {author} {\bibfnamefont {S.}~\bibnamefont
  {Bhadury}}, \bibinfo {author} {\bibfnamefont {W.}~\bibnamefont {Florkowski}},
  \bibinfo {author} {\bibfnamefont {A.}~\bibnamefont {Jaiswal}}, \bibinfo
  {author} {\bibfnamefont {A.}~\bibnamefont {Kumar}},\ and\ \bibinfo {author}
  {\bibfnamefont {R.}~\bibnamefont {Ryblewski}},\ }\bibfield  {title} {\bibinfo
  {title} {{Relativistic Spin Magnetohydrodynamics}},\ }\href
  {https://doi.org/10.1103/PhysRevLett.129.192301} {\bibfield  {journal}
  {\bibinfo  {journal} {Phys. Rev. Lett.}\ }\textbf {\bibinfo {volume} {129}},\
  \bibinfo {pages} {192301} (\bibinfo {year} {2022})},\ \Eprint
  {https://arxiv.org/abs/2204.01357} {arXiv:2204.01357 [nucl-th]} \BibitemShut
  {NoStop}%
\bibitem [{\citenamefont {Acharya}\ \emph {et~al.}(2020)\citenamefont {Acharya}
  \emph {et~al.}}]{ALICE:2019aid}%
  \BibitemOpen
  \bibfield  {author} {\bibinfo {author} {\bibfnamefont {S.}~\bibnamefont
  {Acharya}} \emph {et~al.} (\bibinfo {collaboration} {ALICE}),\ }\bibfield
  {title} {\bibinfo {title} {{Evidence of Spin-Orbital Angular Momentum
  Interactions in Relativistic Heavy-Ion Collisions}},\ }\href
  {https://doi.org/10.1103/PhysRevLett.125.012301} {\bibfield  {journal}
  {\bibinfo  {journal} {Phys. Rev. Lett.}\ }\textbf {\bibinfo {volume} {125}},\
  \bibinfo {pages} {012301} (\bibinfo {year} {2020})},\ \Eprint
  {https://arxiv.org/abs/1910.14408} {arXiv:1910.14408 [nucl-ex]} \BibitemShut
  {NoStop}%
\bibitem [{\citenamefont {Abdallah}\ \emph {et~al.}(2023)\citenamefont
  {Abdallah} \emph {et~al.}}]{STAR:2022fan}%
  \BibitemOpen
  \bibfield  {author} {\bibinfo {author} {\bibfnamefont {M.~S.}\ \bibnamefont
  {Abdallah}} \emph {et~al.} (\bibinfo {collaboration} {STAR}),\ }\bibfield
  {title} {\bibinfo {title} {{Pattern of global spin alignment of
  \ensuremath{\phi} and K$^{*0}$ mesons in heavy-ion collisions}},\ }\href
  {https://doi.org/10.1038/s41586-022-05557-5} {\bibfield  {journal} {\bibinfo
  {journal} {Nature}\ }\textbf {\bibinfo {volume} {614}},\ \bibinfo {pages}
  {244} (\bibinfo {year} {2023})},\ \Eprint {https://arxiv.org/abs/2204.02302}
  {arXiv:2204.02302 [hep-ph]} \BibitemShut {NoStop}%
\bibitem [{\citenamefont {Sheng}\ \emph
  {et~al.}(2023{\natexlab{a}})\citenamefont {Sheng}, \citenamefont {Oliva},
  \citenamefont {Liang}, \citenamefont {Wang},\ and\ \citenamefont
  {Wang}}]{Sheng:2022wsy}%
  \BibitemOpen
  \bibfield  {author} {\bibinfo {author} {\bibfnamefont {X.-L.}\ \bibnamefont
  {Sheng}}, \bibinfo {author} {\bibfnamefont {L.}~\bibnamefont {Oliva}},
  \bibinfo {author} {\bibfnamefont {Z.-T.}\ \bibnamefont {Liang}}, \bibinfo
  {author} {\bibfnamefont {Q.}~\bibnamefont {Wang}},\ and\ \bibinfo {author}
  {\bibfnamefont {X.-N.}\ \bibnamefont {Wang}},\ }\bibfield  {title} {\bibinfo
  {title} {{Spin Alignment of Vector Mesons in Heavy-Ion Collisions}},\ }\href
  {https://doi.org/10.1103/PhysRevLett.131.042304} {\bibfield  {journal}
  {\bibinfo  {journal} {Phys. Rev. Lett.}\ }\textbf {\bibinfo {volume} {131}},\
  \bibinfo {pages} {042304} (\bibinfo {year} {2023}{\natexlab{a}})},\ \Eprint
  {https://arxiv.org/abs/2205.15689} {arXiv:2205.15689 [nucl-th]} \BibitemShut
  {NoStop}%
\bibitem [{\citenamefont {Sheng}\ \emph {et~al.}(2024)\citenamefont {Sheng},
  \citenamefont {Oliva}, \citenamefont {Liang}, \citenamefont {Wang},\ and\
  \citenamefont {Wang}}]{Sheng:2022ffb}%
  \BibitemOpen
  \bibfield  {author} {\bibinfo {author} {\bibfnamefont {X.-L.}\ \bibnamefont
  {Sheng}}, \bibinfo {author} {\bibfnamefont {L.}~\bibnamefont {Oliva}},
  \bibinfo {author} {\bibfnamefont {Z.-T.}\ \bibnamefont {Liang}}, \bibinfo
  {author} {\bibfnamefont {Q.}~\bibnamefont {Wang}},\ and\ \bibinfo {author}
  {\bibfnamefont {X.-N.}\ \bibnamefont {Wang}},\ }\bibfield  {title} {\bibinfo
  {title} {{Relativistic spin dynamics for vector mesons}},\ }\href
  {https://doi.org/10.1103/PhysRevD.109.036004} {\bibfield  {journal} {\bibinfo
   {journal} {Phys. Rev. D}\ }\textbf {\bibinfo {volume} {109}},\ \bibinfo
  {pages} {036004} (\bibinfo {year} {2024})},\ \Eprint
  {https://arxiv.org/abs/2206.05868} {arXiv:2206.05868 [hep-ph]} \BibitemShut
  {NoStop}%
\bibitem [{\citenamefont {Sheng}\ \emph
  {et~al.}(2023{\natexlab{b}})\citenamefont {Sheng}, \citenamefont {Pu},\ and\
  \citenamefont {Wang}}]{Sheng:2023urn}%
  \BibitemOpen
  \bibfield  {author} {\bibinfo {author} {\bibfnamefont {X.-L.}\ \bibnamefont
  {Sheng}}, \bibinfo {author} {\bibfnamefont {S.}~\bibnamefont {Pu}},\ and\
  \bibinfo {author} {\bibfnamefont {Q.}~\bibnamefont {Wang}},\ }\bibfield
  {title} {\bibinfo {title} {{Momentum dependence of the spin alignment of the
  \ensuremath{\phi} meson}},\ }\href
  {https://doi.org/10.1103/PhysRevC.108.054902} {\bibfield  {journal} {\bibinfo
   {journal} {Phys. Rev. C}\ }\textbf {\bibinfo {volume} {108}},\ \bibinfo
  {pages} {054902} (\bibinfo {year} {2023}{\natexlab{b}})},\ \Eprint
  {https://arxiv.org/abs/2308.14038} {arXiv:2308.14038 [nucl-th]} \BibitemShut
  {NoStop}%
\bibitem [{\citenamefont {Kumar}\ \emph
  {et~al.}(2023{\natexlab{a}})\citenamefont {Kumar}, \citenamefont {M\"uller},\
  and\ \citenamefont {Yang}}]{Kumar:2022ylt}%
  \BibitemOpen
  \bibfield  {author} {\bibinfo {author} {\bibfnamefont {A.}~\bibnamefont
  {Kumar}}, \bibinfo {author} {\bibfnamefont {B.}~\bibnamefont {M\"uller}},\
  and\ \bibinfo {author} {\bibfnamefont {D.-L.}\ \bibnamefont {Yang}},\
  }\bibfield  {title} {\bibinfo {title} {{Spin polarization and correlation of
  quarks from the glasma}},\ }\href
  {https://doi.org/10.1103/PhysRevD.107.076025} {\bibfield  {journal} {\bibinfo
   {journal} {Phys. Rev. D}\ }\textbf {\bibinfo {volume} {107}},\ \bibinfo
  {pages} {076025} (\bibinfo {year} {2023}{\natexlab{a}})},\ \Eprint
  {https://arxiv.org/abs/2212.13354} {arXiv:2212.13354 [nucl-th]} \BibitemShut
  {NoStop}%
\bibitem [{\citenamefont {Kumar}\ \emph
  {et~al.}(2023{\natexlab{b}})\citenamefont {Kumar}, \citenamefont {M\"uller},\
  and\ \citenamefont {Yang}}]{Kumar:2023ghs}%
  \BibitemOpen
  \bibfield  {author} {\bibinfo {author} {\bibfnamefont {A.}~\bibnamefont
  {Kumar}}, \bibinfo {author} {\bibfnamefont {B.}~\bibnamefont {M\"uller}},\
  and\ \bibinfo {author} {\bibfnamefont {D.-L.}\ \bibnamefont {Yang}},\
  }\bibfield  {title} {\bibinfo {title} {{Spin alignment of vector mesons by
  glasma fields}},\ }\href {https://doi.org/10.1103/PhysRevD.108.016020}
  {\bibfield  {journal} {\bibinfo  {journal} {Phys. Rev. D}\ }\textbf {\bibinfo
  {volume} {108}},\ \bibinfo {pages} {016020} (\bibinfo {year}
  {2023}{\natexlab{b}})},\ \Eprint {https://arxiv.org/abs/2304.04181}
  {arXiv:2304.04181 [nucl-th]} \BibitemShut {NoStop}%
\bibitem [{\citenamefont {Yang}(2025)}]{Yang:2024qpy}%
  \BibitemOpen
  \bibfield  {author} {\bibinfo {author} {\bibfnamefont {D.-L.}\ \bibnamefont
  {Yang}},\ }\bibfield  {title} {\bibinfo {title} {{Transverse and longitudinal
  spin alignment from color fields in heavy ion collisions}},\ }\href
  {https://doi.org/10.1103/PhysRevD.111.056005} {\bibfield  {journal} {\bibinfo
   {journal} {Phys. Rev. D}\ }\textbf {\bibinfo {volume} {111}},\ \bibinfo
  {pages} {056005} (\bibinfo {year} {2025})},\ \Eprint
  {https://arxiv.org/abs/2411.14822} {arXiv:2411.14822 [nucl-th]} \BibitemShut
  {NoStop}%
\bibitem [{\citenamefont {M{\"u}ller}\ and\ \citenamefont
  {Yang}(2022)}]{Muller:2021hpe}%
  \BibitemOpen
  \bibfield  {author} {\bibinfo {author} {\bibfnamefont {B.}~\bibnamefont
  {M{\"u}ller}}\ and\ \bibinfo {author} {\bibfnamefont {D.-L.}\ \bibnamefont
  {Yang}},\ }\bibfield  {title} {\bibinfo {title} {{Anomalous spin polarization
  from turbulent color fields}},\ }\href
  {https://doi.org/10.1103/PhysRevD.105.L011901} {\bibfield  {journal}
  {\bibinfo  {journal} {Phys. Rev. D}\ }\textbf {\bibinfo {volume} {105}},\
  \bibinfo {pages} {L011901} (\bibinfo {year} {2022})},\ \Eprint
  {https://arxiv.org/abs/2110.15630} {arXiv:2110.15630 [nucl-th]} \BibitemShut
  {NoStop}%
\bibitem [{\citenamefont {Yang}(2022)}]{Yang:2021fea}%
  \BibitemOpen
  \bibfield  {author} {\bibinfo {author} {\bibfnamefont {D.-L.}\ \bibnamefont
  {Yang}},\ }\bibfield  {title} {\bibinfo {title} {{Quantum kinetic theory for
  spin transport of quarks with background chromo-electromagnetic fields}},\
  }\href {https://doi.org/10.1007/JHEP06(2022)140} {\bibfield  {journal}
  {\bibinfo  {journal} {JHEP}\ }\textbf {\bibinfo {volume} {06}},\ \bibinfo
  {pages} {140}},\ \Eprint {https://arxiv.org/abs/2112.14392} {arXiv:2112.14392
  [hep-ph]} \BibitemShut {NoStop}%
\bibitem [{\citenamefont {Carrington}\ \emph {et~al.}(2024)\citenamefont
  {Carrington}, \citenamefont {Mrowczynski},\ and\ \citenamefont
  {Ollitrault}}]{Carrington:2024utf}%
  \BibitemOpen
  \bibfield  {author} {\bibinfo {author} {\bibfnamefont {M.~E.}\ \bibnamefont
  {Carrington}}, \bibinfo {author} {\bibfnamefont {S.}~\bibnamefont
  {Mrowczynski}},\ and\ \bibinfo {author} {\bibfnamefont {J.-Y.}\ \bibnamefont
  {Ollitrault}},\ }\bibfield  {title} {\bibinfo {title} {{Hydrodynamic-like
  behavior of glasma}},\ }\href {https://doi.org/10.1103/PhysRevC.110.054903}
  {\bibfield  {journal} {\bibinfo  {journal} {Phys. Rev. C}\ }\textbf {\bibinfo
  {volume} {110}},\ \bibinfo {pages} {054903} (\bibinfo {year} {2024})},\
  \Eprint {https://arxiv.org/abs/2406.14463} {arXiv:2406.14463 [nucl-th]}
  \BibitemShut {NoStop}%
\bibitem [{\citenamefont {McLerran}\ and\ \citenamefont
  {Venugopalan}(1994{\natexlab{a}})}]{McLerran:1993ni}%
  \BibitemOpen
  \bibfield  {author} {\bibinfo {author} {\bibfnamefont {L.~D.}\ \bibnamefont
  {McLerran}}\ and\ \bibinfo {author} {\bibfnamefont {R.}~\bibnamefont
  {Venugopalan}},\ }\bibfield  {title} {\bibinfo {title} {{Computing quark and
  gluon distribution functions for very large nuclei}},\ }\href
  {https://doi.org/10.1103/PhysRevD.49.2233} {\bibfield  {journal} {\bibinfo
  {journal} {Phys. Rev. D}\ }\textbf {\bibinfo {volume} {49}},\ \bibinfo
  {pages} {2233} (\bibinfo {year} {1994}{\natexlab{a}})},\ \Eprint
  {https://arxiv.org/abs/hep-ph/9309289} {arXiv:hep-ph/9309289} \BibitemShut
  {NoStop}%
\bibitem [{\citenamefont {McLerran}\ and\ \citenamefont
  {Venugopalan}(1994{\natexlab{b}})}]{McLerran:1993ka}%
  \BibitemOpen
  \bibfield  {author} {\bibinfo {author} {\bibfnamefont {L.~D.}\ \bibnamefont
  {McLerran}}\ and\ \bibinfo {author} {\bibfnamefont {R.}~\bibnamefont
  {Venugopalan}},\ }\bibfield  {title} {\bibinfo {title} {{Gluon distribution
  functions for very large nuclei at small transverse momentum}},\ }\href
  {https://doi.org/10.1103/PhysRevD.49.3352} {\bibfield  {journal} {\bibinfo
  {journal} {Phys. Rev. D}\ }\textbf {\bibinfo {volume} {49}},\ \bibinfo
  {pages} {3352} (\bibinfo {year} {1994}{\natexlab{b}})},\ \Eprint
  {https://arxiv.org/abs/hep-ph/9311205} {arXiv:hep-ph/9311205} \BibitemShut
  {NoStop}%
\bibitem [{\citenamefont {McLerran}\ and\ \citenamefont
  {Venugopalan}(1994{\natexlab{c}})}]{McLerran:1994vd}%
  \BibitemOpen
  \bibfield  {author} {\bibinfo {author} {\bibfnamefont {L.~D.}\ \bibnamefont
  {McLerran}}\ and\ \bibinfo {author} {\bibfnamefont {R.}~\bibnamefont
  {Venugopalan}},\ }\bibfield  {title} {\bibinfo {title} {{Green's functions in
  the color field of a large nucleus}},\ }\href
  {https://doi.org/10.1103/PhysRevD.50.2225} {\bibfield  {journal} {\bibinfo
  {journal} {Phys. Rev. D}\ }\textbf {\bibinfo {volume} {50}},\ \bibinfo
  {pages} {2225} (\bibinfo {year} {1994}{\natexlab{c}})},\ \Eprint
  {https://arxiv.org/abs/hep-ph/9402335} {arXiv:hep-ph/9402335} \BibitemShut
  {NoStop}%
\bibitem [{\citenamefont {Gelis}\ \emph {et~al.}(2010)\citenamefont {Gelis},
  \citenamefont {Iancu}, \citenamefont {Jalilian-Marian},\ and\ \citenamefont
  {Venugopalan}}]{Gelis:2010nm}%
  \BibitemOpen
  \bibfield  {author} {\bibinfo {author} {\bibfnamefont {F.}~\bibnamefont
  {Gelis}}, \bibinfo {author} {\bibfnamefont {E.}~\bibnamefont {Iancu}},
  \bibinfo {author} {\bibfnamefont {J.}~\bibnamefont {Jalilian-Marian}},\ and\
  \bibinfo {author} {\bibfnamefont {R.}~\bibnamefont {Venugopalan}},\
  }\bibfield  {title} {\bibinfo {title} {{The Color Glass Condensate}},\ }\href
  {https://doi.org/10.1146/annurev.nucl.010909.083629} {\bibfield  {journal}
  {\bibinfo  {journal} {Ann. Rev. Nucl. Part. Sci.}\ }\textbf {\bibinfo
  {volume} {60}},\ \bibinfo {pages} {463} (\bibinfo {year} {2010})},\ \Eprint
  {https://arxiv.org/abs/1002.0333} {arXiv:1002.0333 [hep-ph]} \BibitemShut
  {NoStop}%
\bibitem [{\citenamefont {Albacete}\ and\ \citenamefont
  {Marquet}(2014)}]{Albacete:2014fwa}%
  \BibitemOpen
  \bibfield  {author} {\bibinfo {author} {\bibfnamefont {J.~L.}\ \bibnamefont
  {Albacete}}\ and\ \bibinfo {author} {\bibfnamefont {C.}~\bibnamefont
  {Marquet}},\ }\bibfield  {title} {\bibinfo {title} {{Gluon saturation and
  initial conditions for relativistic heavy ion collisions}},\ }\href
  {https://doi.org/10.1016/j.ppnp.2014.01.004} {\bibfield  {journal} {\bibinfo
  {journal} {Prog. Part. Nucl. Phys.}\ }\textbf {\bibinfo {volume} {76}},\
  \bibinfo {pages} {1} (\bibinfo {year} {2014})},\ \Eprint
  {https://arxiv.org/abs/1401.4866} {arXiv:1401.4866 [hep-ph]} \BibitemShut
  {NoStop}%
\bibitem [{\citenamefont {Lappi}\ and\ \citenamefont
  {McLerran}(2006)}]{Lappi:2006fp}%
  \BibitemOpen
  \bibfield  {author} {\bibinfo {author} {\bibfnamefont {T.}~\bibnamefont
  {Lappi}}\ and\ \bibinfo {author} {\bibfnamefont {L.}~\bibnamefont
  {McLerran}},\ }\bibfield  {title} {\bibinfo {title} {{Some features of the
  glasma}},\ }\href {https://doi.org/10.1016/j.nuclphysa.2006.04.001}
  {\bibfield  {journal} {\bibinfo  {journal} {Nucl. Phys. A}\ }\textbf
  {\bibinfo {volume} {772}},\ \bibinfo {pages} {200} (\bibinfo {year}
  {2006})},\ \Eprint {https://arxiv.org/abs/hep-ph/0602189}
  {arXiv:hep-ph/0602189} \BibitemShut {NoStop}%
\bibitem [{\citenamefont {Lappi}(2006)}]{Lappi:2006hq}%
  \BibitemOpen
  \bibfield  {author} {\bibinfo {author} {\bibfnamefont {T.}~\bibnamefont
  {Lappi}},\ }\bibfield  {title} {\bibinfo {title} {{Energy density of the
  glasma}},\ }\href {https://doi.org/10.1016/j.physletb.2006.10.017} {\bibfield
   {journal} {\bibinfo  {journal} {Phys. Lett. B}\ }\textbf {\bibinfo {volume}
  {643}},\ \bibinfo {pages} {11} (\bibinfo {year} {2006})},\ \Eprint
  {https://arxiv.org/abs/hep-ph/0606207} {arXiv:hep-ph/0606207} \BibitemShut
  {NoStop}%
\bibitem [{\citenamefont {Schenke}\ \emph {et~al.}(2010)\citenamefont
  {Schenke}, \citenamefont {Jeon},\ and\ \citenamefont
  {Gale}}]{Schenke:2010nt}%
  \BibitemOpen
  \bibfield  {author} {\bibinfo {author} {\bibfnamefont {B.}~\bibnamefont
  {Schenke}}, \bibinfo {author} {\bibfnamefont {S.}~\bibnamefont {Jeon}},\ and\
  \bibinfo {author} {\bibfnamefont {C.}~\bibnamefont {Gale}},\ }\bibfield
  {title} {\bibinfo {title} {{(3+1)D hydrodynamic simulation of relativistic
  heavy-ion collisions}},\ }\href {https://doi.org/10.1103/PhysRevC.82.014903}
  {\bibfield  {journal} {\bibinfo  {journal} {Phys. Rev. C}\ }\textbf {\bibinfo
  {volume} {82}},\ \bibinfo {pages} {014903} (\bibinfo {year} {2010})},\
  \Eprint {https://arxiv.org/abs/1004.1408} {arXiv:1004.1408 [hep-ph]}
  \BibitemShut {NoStop}%
\bibitem [{\citenamefont {Kumar}\ \emph {et~al.}(2024)\citenamefont {Kumar},
  \citenamefont {Yang},\ and\ \citenamefont {Gubler}}]{Kumar:2023ojl}%
  \BibitemOpen
  \bibfield  {author} {\bibinfo {author} {\bibfnamefont {A.}~\bibnamefont
  {Kumar}}, \bibinfo {author} {\bibfnamefont {D.-L.}\ \bibnamefont {Yang}},\
  and\ \bibinfo {author} {\bibfnamefont {P.}~\bibnamefont {Gubler}},\
  }\bibfield  {title} {\bibinfo {title} {{Spin alignment of vector mesons by
  second-order hydrodynamic gradients}},\ }\href
  {https://doi.org/10.1103/PhysRevD.109.054038} {\bibfield  {journal} {\bibinfo
   {journal} {Phys. Rev. D}\ }\textbf {\bibinfo {volume} {109}},\ \bibinfo
  {pages} {054038} (\bibinfo {year} {2024})},\ \Eprint
  {https://arxiv.org/abs/2312.16900} {arXiv:2312.16900 [nucl-th]} \BibitemShut
  {NoStop}%
\bibitem [{\citenamefont {Chen}\ and\ \citenamefont
  {Fries}(2013)}]{Chen:2013ksa}%
  \BibitemOpen
  \bibfield  {author} {\bibinfo {author} {\bibfnamefont {G.}~\bibnamefont
  {Chen}}\ and\ \bibinfo {author} {\bibfnamefont {R.~J.}\ \bibnamefont
  {Fries}},\ }\bibfield  {title} {\bibinfo {title} {{Global Flow of Glasma in
  High Energy Nuclear Collisions}},\ }\href
  {https://doi.org/10.1016/j.physletb.2013.05.031} {\bibfield  {journal}
  {\bibinfo  {journal} {Phys. Lett. B}\ }\textbf {\bibinfo {volume} {723}},\
  \bibinfo {pages} {417} (\bibinfo {year} {2013})},\ \Eprint
  {https://arxiv.org/abs/1303.2360} {arXiv:1303.2360 [nucl-th]} \BibitemShut
  {NoStop}%
\bibitem [{\citenamefont {Chen}\ \emph {et~al.}(2015)\citenamefont {Chen},
  \citenamefont {Fries}, \citenamefont {Kapusta},\ and\ \citenamefont
  {Li}}]{Chen:2015wia}%
  \BibitemOpen
  \bibfield  {author} {\bibinfo {author} {\bibfnamefont {G.}~\bibnamefont
  {Chen}}, \bibinfo {author} {\bibfnamefont {R.~J.}\ \bibnamefont {Fries}},
  \bibinfo {author} {\bibfnamefont {J.~I.}\ \bibnamefont {Kapusta}},\ and\
  \bibinfo {author} {\bibfnamefont {Y.}~\bibnamefont {Li}},\ }\bibfield
  {title} {\bibinfo {title} {{Early Time Dynamics of Gluon Fields in High
  Energy Nuclear Collisions}},\ }\href
  {https://doi.org/10.1103/PhysRevC.92.064912} {\bibfield  {journal} {\bibinfo
  {journal} {Phys. Rev. C}\ }\textbf {\bibinfo {volume} {92}},\ \bibinfo
  {pages} {064912} (\bibinfo {year} {2015})},\ \Eprint
  {https://arxiv.org/abs/1507.03524} {arXiv:1507.03524 [nucl-th]} \BibitemShut
  {NoStop}%
\bibitem [{\citenamefont {Golec-Biernat}\ and\ \citenamefont
  {Wusthoff}(1998)}]{Golec-Biernat:1998zce}%
  \BibitemOpen
  \bibfield  {author} {\bibinfo {author} {\bibfnamefont {K.~J.}\ \bibnamefont
  {Golec-Biernat}}\ and\ \bibinfo {author} {\bibfnamefont {M.}~\bibnamefont
  {Wusthoff}},\ }\bibfield  {title} {\bibinfo {title} {{Saturation effects in
  deep inelastic scattering at low Q**2 and its implications on diffraction}},\
  }\href {https://doi.org/10.1103/PhysRevD.59.014017} {\bibfield  {journal}
  {\bibinfo  {journal} {Phys. Rev. D}\ }\textbf {\bibinfo {volume} {59}},\
  \bibinfo {pages} {014017} (\bibinfo {year} {1998})},\ \Eprint
  {https://arxiv.org/abs/hep-ph/9807513} {arXiv:hep-ph/9807513} \BibitemShut
  {NoStop}%
\bibitem [{\citenamefont {Guerrero-Rodr\'\i{}guez}\ and\ \citenamefont
  {Lappi}(2021)}]{Guerrero-Rodriguez:2021ask}%
  \BibitemOpen
  \bibfield  {author} {\bibinfo {author} {\bibfnamefont {P.}~\bibnamefont
  {Guerrero-Rodr\'\i{}guez}}\ and\ \bibinfo {author} {\bibfnamefont
  {T.}~\bibnamefont {Lappi}},\ }\bibfield  {title} {\bibinfo {title}
  {{Evolution of initial stage fluctuations in the glasma}},\ }\href
  {https://doi.org/10.1103/PhysRevD.104.014011} {\bibfield  {journal} {\bibinfo
   {journal} {Phys. Rev. D}\ }\textbf {\bibinfo {volume} {104}},\ \bibinfo
  {pages} {014011} (\bibinfo {year} {2021})},\ \Eprint
  {https://arxiv.org/abs/2102.09993} {arXiv:2102.09993 [hep-ph]} \BibitemShut
  {NoStop}%
\bibitem [{\citenamefont {Schwinger}(1951)}]{Schwinger:1951nm}%
  \BibitemOpen
  \bibfield  {author} {\bibinfo {author} {\bibfnamefont {J.~S.}\ \bibnamefont
  {Schwinger}},\ }\bibfield  {title} {\bibinfo {title} {{On gauge invariance
  and vacuum polarization}},\ }\href {https://doi.org/10.1103/PhysRev.82.664}
  {\bibfield  {journal} {\bibinfo  {journal} {Phys. Rev.}\ }\textbf {\bibinfo
  {volume} {82}},\ \bibinfo {pages} {664} (\bibinfo {year} {1951})}\BibitemShut
  {NoStop}%
\bibitem [{\citenamefont {Glendenning}\ and\ \citenamefont
  {Matsui}(1983)}]{Glendenning:1983qq}%
  \BibitemOpen
  \bibfield  {author} {\bibinfo {author} {\bibfnamefont {N.~K.}\ \bibnamefont
  {Glendenning}}\ and\ \bibinfo {author} {\bibfnamefont {T.}~\bibnamefont
  {Matsui}},\ }\bibfield  {title} {\bibinfo {title} {{CREATION OF ANTI-Q Q PAIR
  IN A CHROMOELECTRIC FLUX TUBE}},\ }\href
  {https://doi.org/10.1103/PhysRevD.28.2890} {\bibfield  {journal} {\bibinfo
  {journal} {Phys. Rev. D}\ }\textbf {\bibinfo {volume} {28}},\ \bibinfo
  {pages} {2890} (\bibinfo {year} {1983})}\BibitemShut {NoStop}%
\bibitem [{\citenamefont {Kerman}\ \emph {et~al.}(1986)\citenamefont {Kerman},
  \citenamefont {Matsui},\ and\ \citenamefont {Svetitsky}}]{Kerman:1985tj}%
  \BibitemOpen
  \bibfield  {author} {\bibinfo {author} {\bibfnamefont {A.~K.}\ \bibnamefont
  {Kerman}}, \bibinfo {author} {\bibfnamefont {T.}~\bibnamefont {Matsui}},\
  and\ \bibinfo {author} {\bibfnamefont {B.}~\bibnamefont {Svetitsky}},\
  }\bibfield  {title} {\bibinfo {title} {{Particle Production in the Central
  Rapidity Region of Ultrarelativistic Nuclear Collisions}},\ }\href
  {https://doi.org/10.1103/PhysRevLett.56.219} {\bibfield  {journal} {\bibinfo
  {journal} {Phys. Rev. Lett.}\ }\textbf {\bibinfo {volume} {56}},\ \bibinfo
  {pages} {219} (\bibinfo {year} {1986})}\BibitemShut {NoStop}%
\bibitem [{\citenamefont {Gatoff}\ \emph {et~al.}(1987)\citenamefont {Gatoff},
  \citenamefont {Kerman},\ and\ \citenamefont {Matsui}}]{Gatoff:1987uf}%
  \BibitemOpen
  \bibfield  {author} {\bibinfo {author} {\bibfnamefont {G.}~\bibnamefont
  {Gatoff}}, \bibinfo {author} {\bibfnamefont {A.~K.}\ \bibnamefont {Kerman}},\
  and\ \bibinfo {author} {\bibfnamefont {T.}~\bibnamefont {Matsui}},\
  }\bibfield  {title} {\bibinfo {title} {{The Flux Tube Model for
  Ultrarelativistic Heavy Ion Collisions: Electrohydrodynamics of a Quark Gluon
  Plasma}},\ }\href {https://doi.org/10.1103/PhysRevD.36.114} {\bibfield
  {journal} {\bibinfo  {journal} {Phys. Rev. D}\ }\textbf {\bibinfo {volume}
  {36}},\ \bibinfo {pages} {114} (\bibinfo {year} {1987})}\BibitemShut
  {NoStop}%
\bibitem [{\citenamefont {Tanji}(2010)}]{Tanji:2010eu}%
  \BibitemOpen
  \bibfield  {author} {\bibinfo {author} {\bibfnamefont {N.}~\bibnamefont
  {Tanji}},\ }\bibfield  {title} {\bibinfo {title} {{Quark pair creation in
  color electric fields and effects of magnetic fields}},\ }\href
  {https://doi.org/10.1016/j.aop.2010.03.012} {\bibfield  {journal} {\bibinfo
  {journal} {Annals Phys.}\ }\textbf {\bibinfo {volume} {325}},\ \bibinfo
  {pages} {2018} (\bibinfo {year} {2010})},\ \Eprint
  {https://arxiv.org/abs/1002.3143} {arXiv:1002.3143 [hep-ph]} \BibitemShut
  {NoStop}%
\bibitem [{\citenamefont {Taya}(2017)}]{Taya:2016ovo}%
  \BibitemOpen
  \bibfield  {author} {\bibinfo {author} {\bibfnamefont {H.}~\bibnamefont
  {Taya}},\ }\bibfield  {title} {\bibinfo {title} {{Quark and Gluon Production
  from a Boost-invariantly Expanding Color Electric Field}},\ }\href
  {https://doi.org/10.1103/PhysRevD.96.014033} {\bibfield  {journal} {\bibinfo
  {journal} {Phys. Rev. D}\ }\textbf {\bibinfo {volume} {96}},\ \bibinfo
  {pages} {014033} (\bibinfo {year} {2017})},\ \Eprint
  {https://arxiv.org/abs/1609.06189} {arXiv:1609.06189 [nucl-th]} \BibitemShut
  {NoStop}%
\bibitem [{\citenamefont {Florkowski}\ \emph
  {et~al.}(2019{\natexlab{b}})\citenamefont {Florkowski}, \citenamefont
  {Kumar}, \citenamefont {Ryblewski},\ and\ \citenamefont
  {Mazeliauskas}}]{Florkowski:2019voj}%
  \BibitemOpen
  \bibfield  {author} {\bibinfo {author} {\bibfnamefont {W.}~\bibnamefont
  {Florkowski}}, \bibinfo {author} {\bibfnamefont {A.}~\bibnamefont {Kumar}},
  \bibinfo {author} {\bibfnamefont {R.}~\bibnamefont {Ryblewski}},\ and\
  \bibinfo {author} {\bibfnamefont {A.}~\bibnamefont {Mazeliauskas}},\
  }\bibfield  {title} {\bibinfo {title} {{Longitudinal spin polarization in a
  thermal model}},\ }\href {https://doi.org/10.1103/PhysRevC.100.054907}
  {\bibfield  {journal} {\bibinfo  {journal} {Phys. Rev.}\ }\textbf {\bibinfo
  {volume} {C100}},\ \bibinfo {pages} {054907} (\bibinfo {year}
  {2019}{\natexlab{b}})},\ \Eprint {https://arxiv.org/abs/1904.00002}
  {arXiv:1904.00002 [nucl-th]} \BibitemShut {NoStop}%
\bibitem [{\citenamefont {Broniowski}\ and\ \citenamefont
  {Florkowski}(2001)}]{Broniowski:2001we}%
  \BibitemOpen
  \bibfield  {author} {\bibinfo {author} {\bibfnamefont {W.}~\bibnamefont
  {Broniowski}}\ and\ \bibinfo {author} {\bibfnamefont {W.}~\bibnamefont
  {Florkowski}},\ }\bibfield  {title} {\bibinfo {title} {{Explanation of the
  RHIC p(T) spectra in a thermal model with expansion}},\ }\href
  {https://doi.org/10.1103/PhysRevLett.87.272302} {\bibfield  {journal}
  {\bibinfo  {journal} {Phys. Rev. Lett.}\ }\textbf {\bibinfo {volume} {87}},\
  \bibinfo {pages} {272302} (\bibinfo {year} {2001})},\ \Eprint
  {https://arxiv.org/abs/nucl-th/0106050} {arXiv:nucl-th/0106050} \BibitemShut
  {NoStop}%
\bibitem [{\citenamefont {Kanakubo}\ \emph {et~al.}(2022)\citenamefont
  {Kanakubo}, \citenamefont {Tachibana},\ and\ \citenamefont
  {Hirano}}]{Kanakubo:2021qcw}%
  \BibitemOpen
  \bibfield  {author} {\bibinfo {author} {\bibfnamefont {Y.}~\bibnamefont
  {Kanakubo}}, \bibinfo {author} {\bibfnamefont {Y.}~\bibnamefont
  {Tachibana}},\ and\ \bibinfo {author} {\bibfnamefont {T.}~\bibnamefont
  {Hirano}},\ }\bibfield  {title} {\bibinfo {title} {{Interplay between core
  and corona components in high-energy nuclear collisions}},\ }\href
  {https://doi.org/10.1103/PhysRevC.105.024905} {\bibfield  {journal} {\bibinfo
   {journal} {Phys. Rev. C}\ }\textbf {\bibinfo {volume} {105}},\ \bibinfo
  {pages} {024905} (\bibinfo {year} {2022})},\ \Eprint
  {https://arxiv.org/abs/2108.07943} {arXiv:2108.07943 [nucl-th]} \BibitemShut
  {NoStop}%
\bibitem [{\citenamefont {Arnold}\ \emph {et~al.}(2003)\citenamefont {Arnold},
  \citenamefont {Moore},\ and\ \citenamefont {Yaffe}}]{Arnold:2002zm}%
  \BibitemOpen
  \bibfield  {author} {\bibinfo {author} {\bibfnamefont {P.~B.}\ \bibnamefont
  {Arnold}}, \bibinfo {author} {\bibfnamefont {G.~D.}\ \bibnamefont {Moore}},\
  and\ \bibinfo {author} {\bibfnamefont {L.~G.}\ \bibnamefont {Yaffe}},\
  }\bibfield  {title} {\bibinfo {title} {{Effective kinetic theory for high
  temperature gauge theories}},\ }\href
  {https://doi.org/10.1088/1126-6708/2003/01/030} {\bibfield  {journal}
  {\bibinfo  {journal} {JHEP}\ }\textbf {\bibinfo {volume} {01}},\ \bibinfo
  {pages} {030}},\ \Eprint {https://arxiv.org/abs/hep-ph/0209353}
  {arXiv:hep-ph/0209353} \BibitemShut {NoStop}%
\bibitem [{\citenamefont {Kurkela}\ and\ \citenamefont
  {Zhu}(2015)}]{Kurkela:2015qoa}%
  \BibitemOpen
  \bibfield  {author} {\bibinfo {author} {\bibfnamefont {A.}~\bibnamefont
  {Kurkela}}\ and\ \bibinfo {author} {\bibfnamefont {Y.}~\bibnamefont {Zhu}},\
  }\bibfield  {title} {\bibinfo {title} {{Isotropization and hydrodynamization
  in weakly coupled heavy-ion collisions}},\ }\href
  {https://doi.org/10.1103/PhysRevLett.115.182301} {\bibfield  {journal}
  {\bibinfo  {journal} {Phys. Rev. Lett.}\ }\textbf {\bibinfo {volume} {115}},\
  \bibinfo {pages} {182301} (\bibinfo {year} {2015})},\ \Eprint
  {https://arxiv.org/abs/1506.06647} {arXiv:1506.06647 [hep-ph]} \BibitemShut
  {NoStop}%
\bibitem [{\citenamefont {Kurkela}\ \emph {et~al.}(2019)\citenamefont
  {Kurkela}, \citenamefont {Mazeliauskas}, \citenamefont {Paquet},
  \citenamefont {Schlichting},\ and\ \citenamefont {Teaney}}]{Kurkela:2018wud}%
  \BibitemOpen
  \bibfield  {author} {\bibinfo {author} {\bibfnamefont {A.}~\bibnamefont
  {Kurkela}}, \bibinfo {author} {\bibfnamefont {A.}~\bibnamefont
  {Mazeliauskas}}, \bibinfo {author} {\bibfnamefont {J.-F.}\ \bibnamefont
  {Paquet}}, \bibinfo {author} {\bibfnamefont {S.}~\bibnamefont
  {Schlichting}},\ and\ \bibinfo {author} {\bibfnamefont {D.}~\bibnamefont
  {Teaney}},\ }\bibfield  {title} {\bibinfo {title} {{Matching the
  Nonequilibrium Initial Stage of Heavy Ion Collisions to Hydrodynamics with
  QCD Kinetic Theory}},\ }\href
  {https://doi.org/10.1103/PhysRevLett.122.122302} {\bibfield  {journal}
  {\bibinfo  {journal} {Phys. Rev. Lett.}\ }\textbf {\bibinfo {volume} {122}},\
  \bibinfo {pages} {122302} (\bibinfo {year} {2019})},\ \Eprint
  {https://arxiv.org/abs/1805.01604} {arXiv:1805.01604 [hep-ph]} \BibitemShut
  {NoStop}%
\bibitem [{\citenamefont {Huang}(2025)}]{Huang:2024ffg}%
  \BibitemOpen
  \bibfield  {author} {\bibinfo {author} {\bibfnamefont {X.-G.}\ \bibnamefont
  {Huang}},\ }\bibfield  {title} {\bibinfo {title} {{An introduction to
  relativistic spin hydrodynamics}},\ }\href
  {https://doi.org/10.1007/s41365-025-01784-3} {\bibfield  {journal} {\bibinfo
  {journal} {Nucl. Sci. Tech.}\ }\textbf {\bibinfo {volume} {36}},\ \bibinfo
  {pages} {208} (\bibinfo {year} {2025})},\ \Eprint
  {https://arxiv.org/abs/2411.11753} {arXiv:2411.11753 [nucl-th]} \BibitemShut
  {NoStop}%
\end{thebibliography}%

\end{document}